\documentclass{article}
\usepackage{graphicx}
\usepackage{epsfig}
\usepackage[utf8]{inputenc}
\usepackage{color}
\usepackage{amssymb}
\usepackage{amsmath}
\usepackage{subfigure}
\usepackage{float}
\usepackage{geometry}
\usepackage{booktabs}
\usepackage{caption}
\usepackage{makecell}
\usepackage{natbib}
\usepackage{authblk}
\setcitestyle{authoryear,round}
\usepackage[colorlinks,
            linkcolor=blue,
            anchorcolor=blue,
            urlcolor=blue,
            citecolor=blue]{hyperref}

\title{CoGANPPIS: A Coevolution-enhanced Global Attention Neural Network for Protein-Protein Interaction Site Prediction}

\author[1]{Jiaxing Guo}
\author[1]{Xuening Zhu}
\author[2,3]{Zixin Hu}
\author[2,*]{Xiaoxi Hu}
\affil[1]{School of Data Science, Fudan University, Shanghai, China.}
\affil[2]{State Key Laboratory of Genetic Engineering and Innovation Center of Genetics and Development, School of Life Sciences, Fudan University, Shanghai, China.}
\affil[3]{Artificial Intelligence Innovation and Incubation Institute, Fudan University, Shanghai, China.}
\affil[*]{Corresponding author. E-mail: xxhu21@m.fudan.edu.cn, Address: No.2005 Songhu Road, Yangpu District, Shanghai 200438, China}

\begin{document}

\maketitle

\begin{abstract}

Protein-protein interactions are of great importance in biochemical processes. Accurate prediction of protein-protein interaction sites (PPIs) is crucial for our understanding of biological mechanism. Although numerous approaches have been developed recently and achieved gratifying results, there are still two limitations: (1) Most existing models have excavated a number of useful input features, but failed to take coevolutionary features into account, which could provide clues for inter-residue relationships; (2) The attention-based models only allocate attention weights for neighboring residues, instead of doing it globally, which may limit the model's prediction performance since some residues being far away from the target residues might also matter.

We propose a coevolution-enhanced global attention neural network, a sequence-based deep learning model for PPIs prediction, called CoGANPPIS. Specifically, CoGANPPIS utilizes three layers in parallel for feature extraction: (1) Local-level representation aggregation layer, which aggregates the neighboring residues' features as the local feature representation; (2) Global-level representation learning layer, which employs a novel coevolution-enhanced global attention mechanism to allocate attention weights to all residues on the same protein sequences; (3) Coevolutionary information learning layer, which applies CNN \& pooling to coevolutionary information to obtain the coevolutionary profile representation. Then, the three outputs are concatenated and passed into several fully connected layers for the final prediction. Extensive experiments on two benchmark datasets have been conducted, demonstrating that our proposed model achieves the state-of-the-art performance.

\end{abstract}

\section{Introduction}

Proteins participate in a variety of biological processes in organisms. They rarely act alone, instead, they usually carry out various functions by interacting with different kinds of molecules, such as DNA, lipids, carbohydrates, and other proteins \citep{branden2012introduction,murray2009harper,ardejani2017using}. The process of establishing physical contacts of high specificity between two or more protein molecules is known as protein-protein interaction, which plays an important role in many biochemical processes including immune response, muscle contraction, and signal transduction. Considering the high practical and research value of PPIs prediction, many approaches have been proposed so far. There are some conventional experimental methods, such as two-hybrid screening, spectrofluorometry, and intragenic complementation, being commonly applied to identify PPIs \citep{westermarck2013identification,terentiev2009dynamic,brettner2012protein,smallwood2002intragenic}. However, these experimental methods suffer from being costly and time-consuming so that more accurate and efficient computational predictors for PPIs are of great value for biologists.

With the rapid development of computer science, a lot of computational approaches, especially ML-based approaches, have been developed, which take protein sequences or structures as input and are known as the sequence-based and the structure-based respectively \citep{hou2016club}. Although Structure-based methods have achieved some promising progress in recent years \citep{gainza2020deciphering,yuan2022structure,huang2023sgppi}, they may cause problems for biological researchers since the number of proteins with available structures is limited. AlphaFold2 has shown promising performance in protein structure prediction, but its effectiveness on some newly-discovered proteins and some exotic proteins still remains to be tested. Moreover, its requirement for computational resources could be too high for most researchers \citep{jumper2021highly}. In contrast, sequence-based methods are more practical since protein sequences are easier to be obtained with the noticeable development of high-throughput techniques. 

Sequence-based methods could be classiﬁed as the partner-speciﬁc and the non partner-speciﬁc \citep{casadio2022machine}, and in this paper we focus on the later one. The partner-speciﬁc sequence-based PPIs prediction aims to identify the interaction residue pairs of two given proteins, which has not been covered in our present work.


Sequence-based methods can be further classified into 2 categories: traditional machine learning approaches and deep learning approaches. The commonly-used traditional machine learning approaches include SVM \citep{yan2004two,wang2006predicting,porollo2007prediction,chen2010sequence,chen2012detection}, Na\"ive Bayes \citep{yan2004two,murakami2010applying}, shallow neural network \citep{ofran2003predicted,ofran2007isis}, random forest \citep{chen2009sequence,northey2018intpred,wang2019protein}, and logistic regression \citep{dhole2014sequence,zhang2019scriber}. However, these methods cannot effectively excavate the interaction between different features and suffer from overfitting.

In recent years, due to the great success of deep learning in many fields such as computer vision, speech recognition and natural language processing, the models based on deep learning have also been used in PPIs prediction. Among these, DeepPPISP, a model based on convolutional neural networks, showed significant improvement in PPIs prediction compared with traditional machine learning models \citep{zeng2020protein}. DELPHI achieved good results through a fine-tuned ensemble neural network combining recurrent neural networks and convolutional neural networks \citep{li2021delphi}. Then, an attention-based convolutional neural network, which we call ACNN, made a better understanding of the local environment of the target residues by giving different attention weights to the neighboring residues \citep{lu2021attention}. HANPPIS improved the performance and interpretability by using a double-layer attention mechanism \citep{tang2021prediction}. Besides, ensnet\_p, an ensemble model combining several neural net architectures, achieved stable and high prediction accuracy \citep{stringer2022pipenn}.

Conventional sequence-based input features can be roughly classified into 4 categories: raw sequence features, evolutionary information, residue physiochemical properties, and predicted structural features \citep{casadio2022machine}. Raw sequence features refer to the features that can be straightly obtained by the protein sequences, the most commonly used of which are amino acid types. Evolutionary information usually refers to the position-specific scoring matrices (PSSM) as well as other conservative scores of the proteins, which are mainly calculated from multiple sequence alignments (MSA) and very informative for protein-related prediction tasks. Residue physiochemical properties (such as residue's charge and polarity) have been applied in many models in recent years, which can be obtained from databases or some specific predictors. Besides, in the absence of protein structure, predicted structural features (such as hydrophobicity, secondary structure, disorder) can also be helpful.

Recently, another sequence-based feature, coevolutionary information based feature, has been applied to another important protein-related problem, protein contact-map prediction, and brings about significant performance improvement \citep{wang2017accurate,hanson2018accurate,li2021deducing}. These features are mainly obtained by direct coupling analysis (DCA) and could quantify the inter-residue coevolutionary relationships. Intuitively, for a residue, its properties and behaviours should be more similar to the residues closely related to it, which inspires us to introduce it into our model.

To evaluate the performance of our model, we compare it with seven other sequence-based models (PSIVER, ISIS, SPRINGS, DELPHI, DeepPPISP, ACNN and ensnet\_p) on two benchmark datasets Dset422 and Dset448. The experimental results show that our model achieves state-of-the-art performance for PPIs prediction. The main contributions of this paper are as follows:

(1) To the best of our knowledge, for non partner-specific PPIs prediction, this is the first time to introduce coevolutionary information as input features into the model. And we verify its usefulness by ablation analysis.

(2) We propose a novel coevolution-enhanced global attention mechanism for global-level representation learning, which allocates attention weights based on a better understanding of the whole protein sequences and the coevolutionary relationships among the residues.


(3) We provide a user-friendly package with feature construction pipelines so that users only need to input the raw protein sequences to obtain the prediction results.

The rest of this paper is organized as follows: Section 2 introduces the datasets, the input features and the architecture of our model. Section 3 presents the experimental results and experiment analysis. In the end, section 4 summarizes this paper.



\section{Materials and Methods}

\subsection{Datasets}

In this study, two benchmark datasets, Dset422 and Dset448, are used in experiments for model comparison. Dset422 consists of Dset72 \citep{murakami2010applying}, Dset186, and Dset164 \citep{singh2014springs}, whose protein sequences are collected from Protein Data Bank \citep{sussman1999bridging}. The protein sequence homology is less than 25\% and if an amino acid has an absolute solvent proximity less than 1 $\text{\r{A}}^{2}$ before and after binding with other proteins, it will be defined as an interaction site \citep{zeng2020protein}. Dset448 is sourced from the BioLip database, where residues are defined as interaction sites if the distance between an atom of this residue and an atom of a given protein-partner is less than 0.5 $\text{\r{A}}$ plus the sum of the Van der Waal's radii of the two atoms \citep{zhang2019scriber}. First, the protein sequences were mapped into Uniprot databases to collect binding residues across different complexes. Then, they were clustered by Blastclust at 25\% similarity, after which one protein was selected from each cluster to ensure that proteins in Dset448 shared similarities less than 25\%. Besides, a dataset of 4392 protein sequences was constructed from PDB in the paper of PIPENN, where interaction sites are defined similar to Dset448 \citep{stringer2022pipenn}. Here we call it Dset4392. In order to reduce the sequence similarities, we utilized Blastclust to cluster all the protein sequences in Dset422, Dset448 and Dset4392 at the similarity of 25\%. We removed those from Dset4392 whose clusters also contained sequences from Dset422 or Dset448. We use the remaining 4331 sequences as the pretraining dataset and call it Dset4331 in this study.

For the three datasets, we randomly divided them into training set (about 83\% of randomly selected proteins), validation set (about 5\% of randomly selected proteins), and test set (the remaining proteins) respectively. Consequently, for Dset422, there are 352 proteins in the training set, 21 proteins in the validation set, and 49 proteins in the test set. For Dset448, there are 373 proteins in the training set, 22 proteins in the validation set, and 53 proteins in the test set. And For Dset4331, there are 3594 proteins in the training set, 216 proteins in the validation set, and 521 proteins in the test set. The statistics of the three datasets are shown in Table \ref{the_length_of_sequences}.

\begin{table}[h]
\centering
\caption{Statistics of the three datasets}
\label{the_length_of_sequences}
\begin{tabular*}{0.8\textwidth}{lccc}
\hline
  & Dset422 & Dset448 & Dset4331 \\
\hline
No. proteins & 422 & 448 & 4331\\
No. residues & 88 040 & 116 500 & 1 263 453\\
No. interaction sites & 13 536 & 15 810 & 149 342\\
No. non-interaction sites & 74 504 & 100 690 & 1 114 111\\
Percentage of interaction sites & 15.37\% & 13.57\% & 11.82\%\\
No. train proteins(residues) & 352(75 398) & 373(100 189) & 3594(1 042 757)\\
No. validate proteins(residues) & 21(3593) & 22(4522) & 216(62 890)\\
No. test proteins(residues) & 49(9049) & 53(11 789) & 521(157 806)\\
\hline
\end{tabular*}
\end{table}

\subsection{Input Features}

The features commonly used in previous research, including raw sequence features, position-specific scoring matrices, and predicted secondary structures, are applied in our model. Besides, we also introduce coevolutionary information into our model.

\subsubsection{Raw sequence features}

In this study, we utilize two raw sequence features, amino acid type and sequence length. Most proteins consist of 20 different amino acids. Hence, we encode each amino acid residue as a 20D one-hot vector representing the amino acid type at this position. Besides, we utilize an integer to represent the length of the sequence for each residue as another raw sequence feature.

\subsubsection{Position-specific scoring matrices}

Position-specific scoring matrices (PSSM) contain evolutionary information, which have been shown effective for PPIs prediction \citep{cheol2010position}. We perform the PSI-BLAST algorithm on each input sequence against NCBI's non-redundant sequence database with three iterations and an E-value threshold of 0.001 to obtain its PSSM, where every amino acid residue on the underlying sequence is encoded as a vector with 20 elements.

\subsubsection{Predicted secondary structures}

NetSurfP-3.0 is a tool for predicting secondary structures from protein sequences \citep{hoie2022netsurfp}. Here we utilize it to predict relative solvent accessibility (RSA), access surface area (ASA), 3-state secondary structure as well as 8-state secondary structure for each residue. Each amino acid residue is encoded as a vector with 13 elements, which represents the predicted RSA, predicted ASA, and the predicted probabilities of being the corresponding secondary structure states at the position.


\subsubsection{Coevolutionary information}

Coevolutionary relationships between amino acid residues refer to the interdependent changes that occur in pairs of residues on the same protein sequences, which help maintain proteins' stability, function, and folding \citep{de2013emerging}. As mentioned earlier, coevolutionary information based features brought about great performance improvement in protein-contact map prediction, inspiring us to apply it in this study.

Direct-coupling analysis (DCA) is one of the main computational approaches to capture proteins' coevolutionary information. The key idea of DCA is to disentangle direct pairwise couplings of each two amino acid residues on the same protein sequences. For each protein, DCA takes its multiple sequence alignments (MSA) as the input, which is obtained by BLASTP, and returns a $N\times N$ matrix, where $N$ refers to the length of the protein sequence. The $(i,j)$ element of the matrix refers to the direct coupling degree of the $i$th residue and the $j$th residue on the protein sequence. The larger the value, the higher the coevolutionary relationship exists between these two residues. In this study, DCA features are used for constructing the attention mechanism and generating coevolutionary profile representation.

There are three usual DCA algorithms, mpDCA \citep{weigt2009identification}, mfDCA \citep{morcos2011direct}, and plmDCA \citep{ekeberg2013improved}. The mpDCA uses a semi-heuristic message-passing approach, and its slow computation speed makes it difficult to be applied to a large-scale dataset. The mfDCA uses a mean-field approach based on the maximum-entropy model, which greatly improves the computational speed. On the basis of mfDCA, the plmDCA applies pseudo-likelihood to the Potts model and achieves higher accuracy than mfDCA. Based on the above comparison, we utilize plmDCA to generate DCA matrices.

\subsection{Model Architecture}

\begin{figure*}
\centering
\includegraphics[width=1\textwidth]{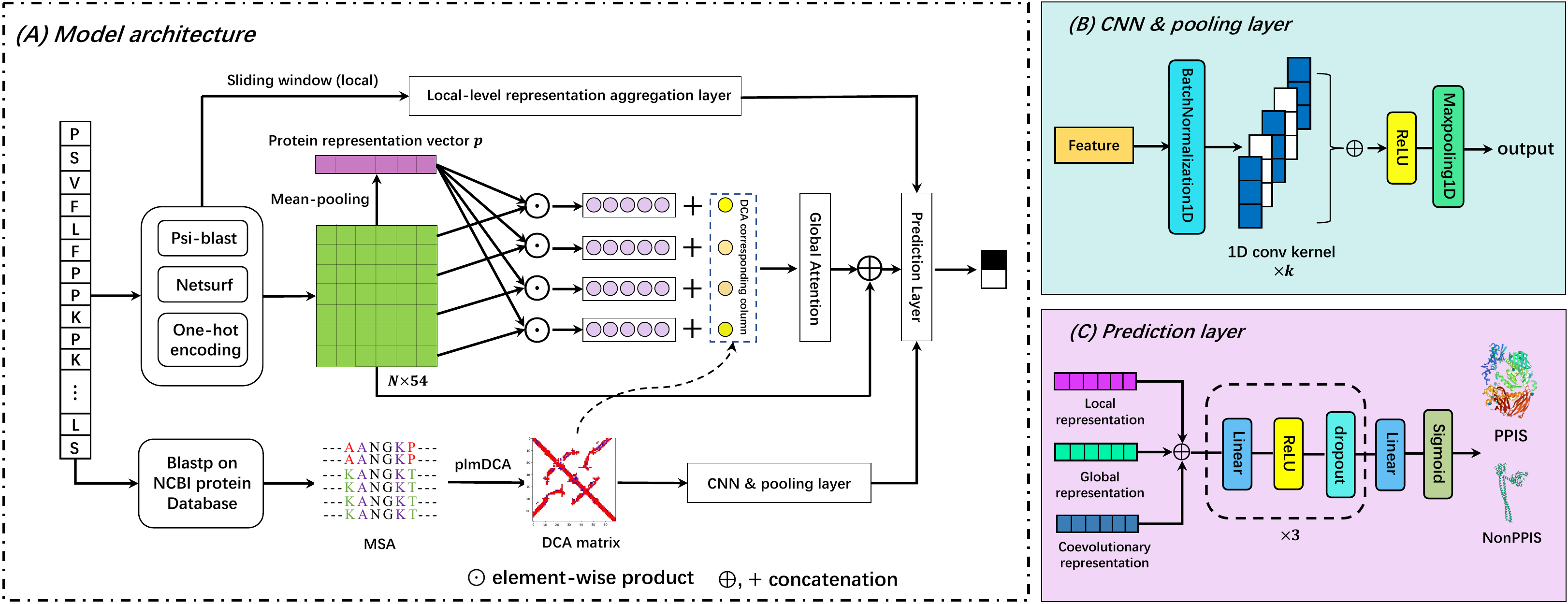}
\caption{The model architecture of CoGANPPIS. (A) An overall illustration of the proposed model. The feature extraction consists of three layers: local-level representation aggregation layer, global-level representation learning layer and coevolutionary information learning layer, whose outputs are passed into the prediction layer for the final prediction. (B) The structure of CNN \& pooling in coevolutionary information learning layer. We use batchnormalization, multiple convolution kernels, activation as well as maxpooling to capture the coevolutionary information from DCA. (C) The structure of the prediction layer. We concatenate the outputs from feature extraction and utilize four fully-connected layers to predict whether the residue is an interaction site or not.}\label{model_fig}
\end{figure*}

Figure \ref{model_fig} is an overview of the proposed framework. Firstly, we extract local features, global features, and coevolutionary features based on local level representation aggregation layer, global level representation learning layer, and coevolutionary information learning layer, respectively. Then we concatenate three feature representation vectors as the output of the feature extraction and pass it into the prediction layer consisting of four fully connected layers for the final prediction about whether the target amino acid residue is an interaction site or not. Now we introduce them in detail.

\subsubsection{Local-level representation aggregation layer}

For each target residue, a sliding window of length $(2n+1)$ is used to aggregate the features of itself and its neighboring $2n$ residues. For the $i$th residue on the protein sequence, we calculate its local feature representation $h_i^{local}$ by concatenating these residues' features.

\subsubsection{Global-level representation learning layer}

It has been shown that the global features of protein sequences are critical for PPIs prediction \citep{zeng2020protein}. Also, the coevolutionary information could quantify the inter-residue coevolutionary relationships \citep{wang2017accurate,hanson2018accurate,li2021deducing}. Hence, we devise a coevolution-enhanced global attention mechanism to distinguish the importance of residues. Assuming that the $i$th residue is the predicting target, all residues on the same sequence are linearly added according to the attention scores,

\begin{equation}
    h_{i}^{p}=\sum_{j=1}^{N}\alpha_{ij}h_{j},
\end{equation}

\noindent{where $h_j$ refers to the PSSM, predicted secondary structure, and raw protein sequence features of residue $j$. $\alpha_{ij}$ refers to the attention score, which estimates the importance weight of residue $j$ on target residue $i$. Intuitively, different residues on the same protein sequence should have different importance for the target residue, and those that match the charateristics of the whole protein and share a close relationship with the target residue should be paid more attention to. Therefore, $\alpha_{ij}$ is calculated as follows:}

\begin{equation}
    \alpha_{ij}={\rm softmax}(\pi(i,j))=\frac{\exp{(\pi(i,j))}}{\sum_{k=1}^{N}\exp(\pi(i,k))},
\end{equation}

\begin{equation}
    \pi(i,j)=q_{1}^{\rm T}{\rm LeakyRelu}([W_{1}(p\odot h_{j})\Vert W_{2}(h_{i}\odot h_{j})]\Vert w_{ij}).
\end{equation}

Here, we use LeakyRelu as the activation function. $\odot$ indicates element-wise product, $\Vert$ indicates concatenation operation, and $w_{ij}\in\mathbb{R}^{1}$ refers to the $(i,j)$ element of DCA matrix of the current protein, which provides us some clues for the coevolutionary relationship between the residue $i$ and residue $j$. $W_1\in\mathbb{R}^{d\times d}$, $W_2\in\mathbb{R}^{d\times d}$ and $q_1\in\mathbb{R}^{2d+1}$ are trainable parameters. $p$ can be seen as the feature representation of the whole protein sequence, which is obtained by mean-pooling on all the residues' features on this protein,

\begin{equation}
    p=\frac{1}{N}\sum_{j=1}^{N}h_j.
\end{equation}

Our approach makes the attention weights between the target residue and other residues dependent on not only the whole protein sequence feature representation but also the coevolutionary relationships between them, suggesting that those residues which match the charateristics of the whole protein and are closely related to the target residue will be attached more importance.

Then we concatenate the global representation of the target residue $h_i^{p}$ and its original feature $h_i$,

\begin{equation}
    h_i^{global} = h_i^{p}\Vert h_i,
\end{equation}

\noindent{where $h_i^{global}$ is the result of the global-level representation learning layer.}

\subsubsection{Coevolutionary information learning layer}

We have introduced coevolutionary information into the attention mechanism to exploit the relationship among residues as above. Now we further utilize the coevolutionary information on a larger scale, i.e, on the whole protein sequence level. Suppose we are predicting the $i$th residue on the protein sequence. First, we take its corresponding column in the DCA matrix as its coevolutionary information. Then we pass it into the CNN \& pooling layer as shown in Figure \ref{model_fig}:

\begin{equation}
    h_{i,k}^{dca}={\rm Relu}({\rm conv1d}^{(k)}({\rm BN}({\rm DCA}[:,i]))), k\in[1,K],
\end{equation}

\begin{equation}
    h_{i}^{dca}=\Vert_{k=1}^{K}h_{i,k}^{dca},
\end{equation}

\noindent{where ${\rm DCA}[:,i]$ is the $i$th column of the DCA matrix of the underlying protein. ${\rm BN}$ refers to the BatchNormalization operation and the 1D convolution operation ${\rm conv1d}$ extracts the normalized coevolutionary features. We use ${\rm Relu}$ as the activation function. Here we use $K$ different convolution kernels for a better extraction of coevolutionary features. Finally, we obtain the coevolutionary representation $h_{i}^{dca}$ by concatenating all $K$ results linearly.}

\subsubsection{Prediction layer}

To predict whether an amino acid residue is an interaction site or not, first we concatenate three former feature extraction results to obtain the final representation:

\begin{equation}
    h_{i}^{pred}=h_i^{local}\Vert h_i^{global}\Vert h_{i}^{dca},
\end{equation}

\noindent{where $h_i^{local}$, $h_i^{global}$ and $h_{i}^{dca}$ are the results of local-level representation aggregation layer, global-level representation learning layer, and coevolutionary information learning layer. $h_{i}^{pred}$ is the final representation of the residue, which will be passed into fully connected layers:}

\begin{equation}
    x^{(t)}={\rm Relu}(W^{(t)}x^{(t-1)}+b^{(t)}), t\in [1,T],
\end{equation}

\noindent{where $x^{(t)}$ and $x^{(t-1)}$ refer to the input vector and output vector of the $t$th fully connected layer, respectively. Here, $x^{(0)}=h_{i}^{pred}$. $W^{(t)}$ denotes the weight matrix and $b^{(t)}$ denotes the bias. Besides, ${\rm ReLU}$ and ${\rm dropout}$ are utilized in each layer except the last one. After the last layer, a Sigmoid function is used to generate the final prediction:}

\begin{equation}
    \hat{y}=\frac{1}{1+e^{-x^{(T)}}},
\end{equation}

\noindent{where $\hat{y}$ denotes the predicted probability of the residue being an interaction site. And $1-\hat{y}$ is the predicted probability of the residue being a non-interaction site.}

\subsection{Evaluation Metrics}

PPIs prediction can be seen as a binary classification problem for identifying whether an amino acid residue is an interaction site or not. Consequently, there could be four types of results based on the residue's true category and predicted category, i.e., true positives (TP), true negatives (TN), false positives (FP), and false negatives (FN). Here, TP and TN refer to the correctly predicted interaction sites and non-interaction sites respectively; FP and FN refer to the incorrectly predicted interaction sites and non-interaction sites respectively.

We select six evaluation metrics to comprehensively evaluate the predictive performance, including area under the precision-recall curve (AUPRC), accuracy (ACC), recall, precision, F-measure (${\rm F_1}$), and Matthews correlation coefficient (MCC). Considering that our dataset is imbalanced with more non-interaction sites than interaction sites, ${\rm F_1}$ and MCC indices deserve more attention. The formulas for calculating these metrics are as follows:

\begin{equation}
{\rm ACC=\frac{TP+TN}{TP+TN+FP+FN}},
\end{equation}
\begin{equation}
{\rm Recall=\frac{TP}{TP+FN}},
\end{equation}
\begin{equation}
{\rm Precision=\frac{TP}{TP+FP}},
\end{equation}
\begin{equation}
{\rm F_1={\rm \frac{2\times Precision\times Recall}{Precision+Recall}}},
\end{equation}
\begin{equation}
{\rm MCC}={\rm \frac{TP\times TN - FP\times FN}{\sqrt{(TP+FP)(TP+FN)(TN+FP)(TN+FN)}}}.
\end{equation}


\section{Results and Discussion}

\subsection{Implementation}

In the feature extraction part, the sliding window length in the local-level representation aggregation layer is set as 7. We use three convolution kernels in the CNN \& pooling layer and the sizes are set as 13, 15 and 17, respectively. In the classification part, we utilized four fully connected layers of 1024, 256, 8, and 1 node, with the first three fully connected layers accompanied by a dropout ratio of 0.1. We use weighted cross-entropy loss as the loss function:

\begin{equation}
    L=-\frac{1}{m}\sum\limits_{i=1}^{m}(wy_ilog(\hat{y}_i)+(1-y_i)log(1-\hat{y}_i)),
\end{equation}

\noindent{where $m$ is the number of training samples. $w$ refers to the weight and is set to $4$. Interaction site is labeled as 1 ($y_i=1$) and non-interaction site is labeled as 0 ($y_i=0$). $\hat{y}_i$ is the predicted probability of being interaction site of the sample $i$. Besides, we utilize Adaptive Momentum (Adam) as the optimizer with a learning rate of 0.0001. The batch size is set to 256. The model is implemented by PyTorch and trained on NVIDIA GTX 1080 Ti.}

Fine tuning is used in this model. Before training on Dset422 and Dset448, we first trained our model on Dset4331. In the epoch of achieving the best performance on Dset4331, the parameters of the model are saved to files. When training on Dset422 and Dset448, we loaded the saved weights from the file and froze the feature extraction weights so that during the process of training, the parameters in feature extraction stayed unchanged. Training data are used only to train the fully connected layers in the prediction layer.

\subsection{Comparison with competing methods}

To evaluate the predictive performance of our model (CoGANPPIS), we compare it with seven popular sequence-based competing methods (PSIVER, ISIS, SPRINGS, DELPHI, DeepPPISP, ACNN and ensnet\_p). Specifically, PSIVER \citep{murakami2010applying} utilizes Na\"ive Bayes Classifier and kernel density estimation method to predict PPIs based on sequence features. ISIS \citep{ofran2007isis} combines predicted structural features with evolutionary information to predict PPIs based on shallow neural networks. SPRINGS utilizes an artificial neural network to generate PPIs predictions \citep{singh2014springs}. DELPHI \citep{li2021delphi} employs a fine-tuned ensemble model by combining several recurrent neural networks and convolutional neural networks. DeepPPISP \citep{zeng2020protein} considers both local contextual and global information and applies a convolutional neural network to predict PPIs. ACNN \citep{lu2021attention} employs a local attention mechanism to make PPIs prediction. And ensnet\_p is an ensemble model combining different neural net models \citep{stringer2022pipenn}. In words, all these traditional approaches do not apply coevolutionary information or global attention mechanism.

\begin{figure}[h]
	\centering  
	\subfigbottomskip=0pt 
	\subfigcapskip=0pt 
	\subfigure[]{
	    \label{dset422_ROC}
		\includegraphics[width=0.34\linewidth]{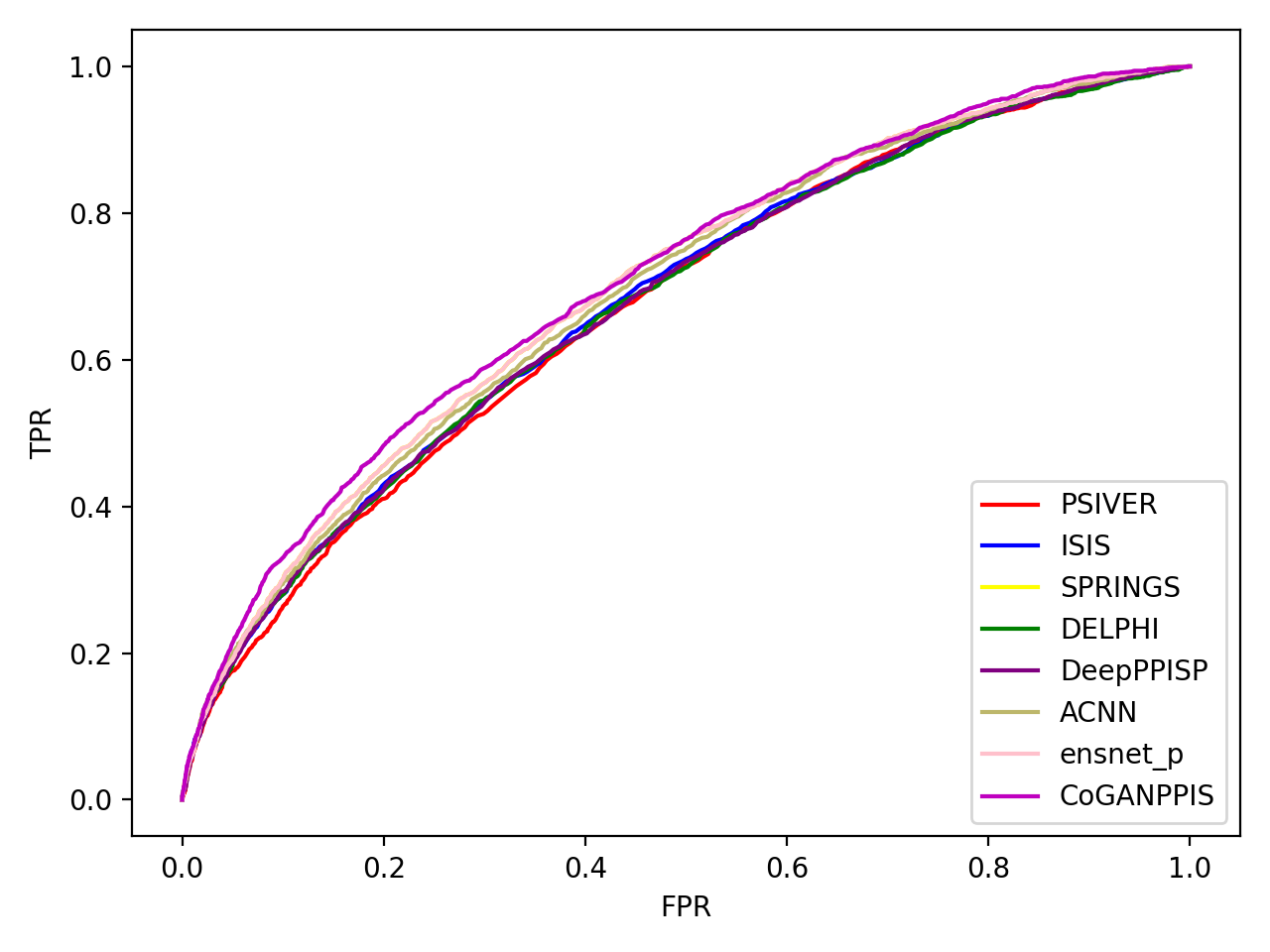}}
	\subfigure[]{
	    \label{dset422_PR}
		\includegraphics[width=0.34\linewidth]{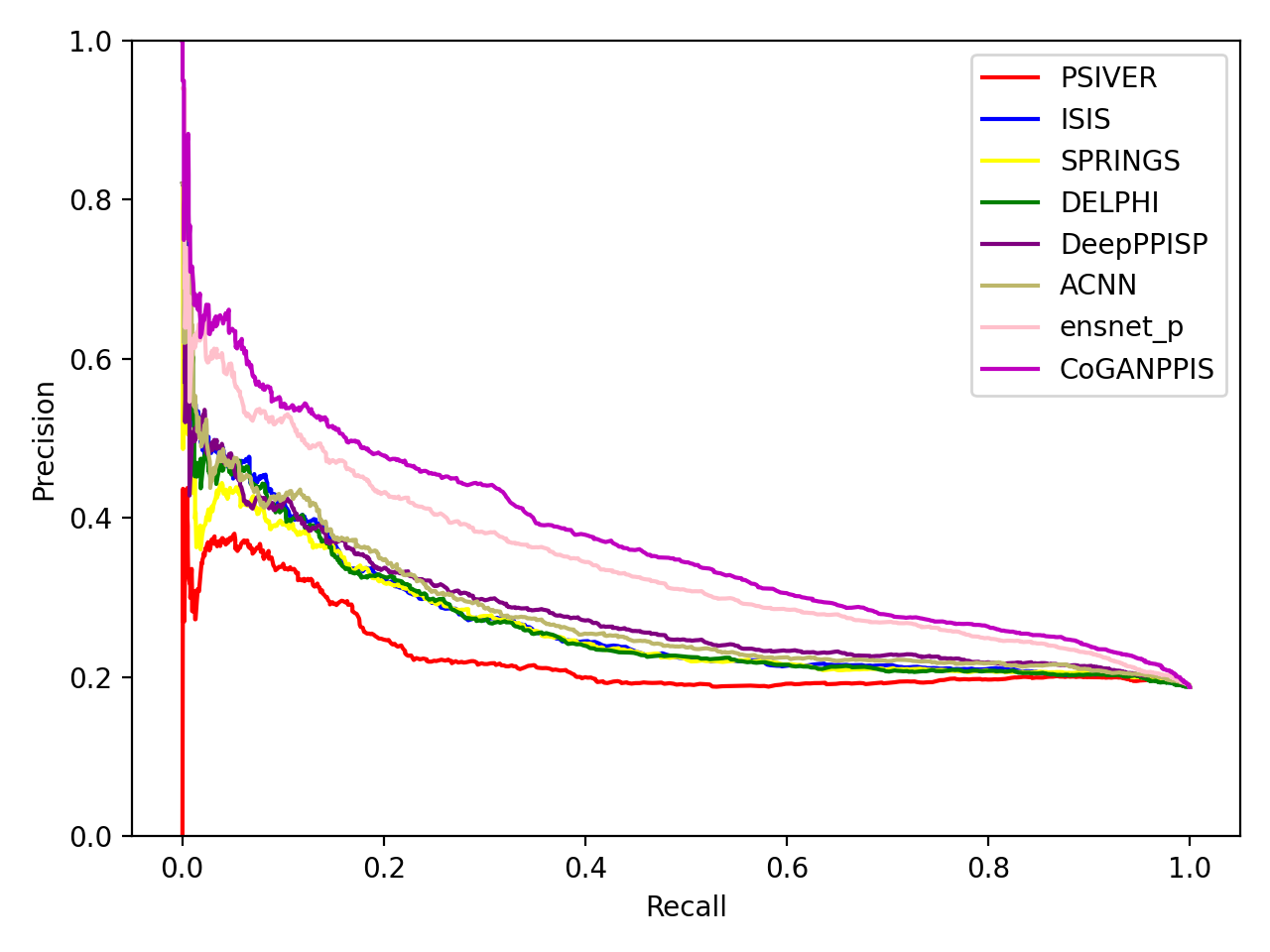}}
    \subfigure[]{
	    \label{dset448_ROC}
		\includegraphics[width=0.34\linewidth]{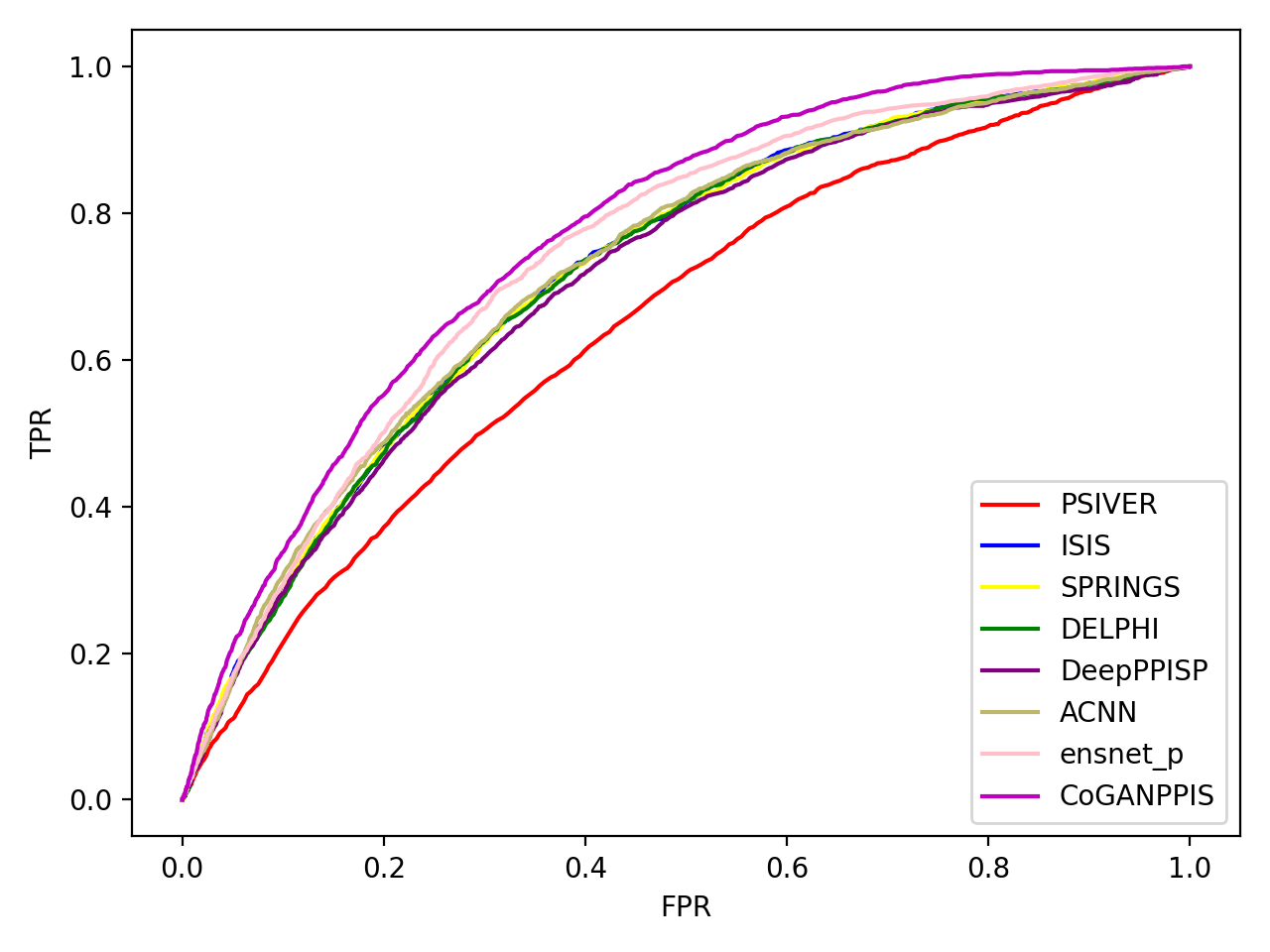}}
    \subfigure[]{
	    \label{new_dset448_PR}
		\includegraphics[width=0.34\linewidth]{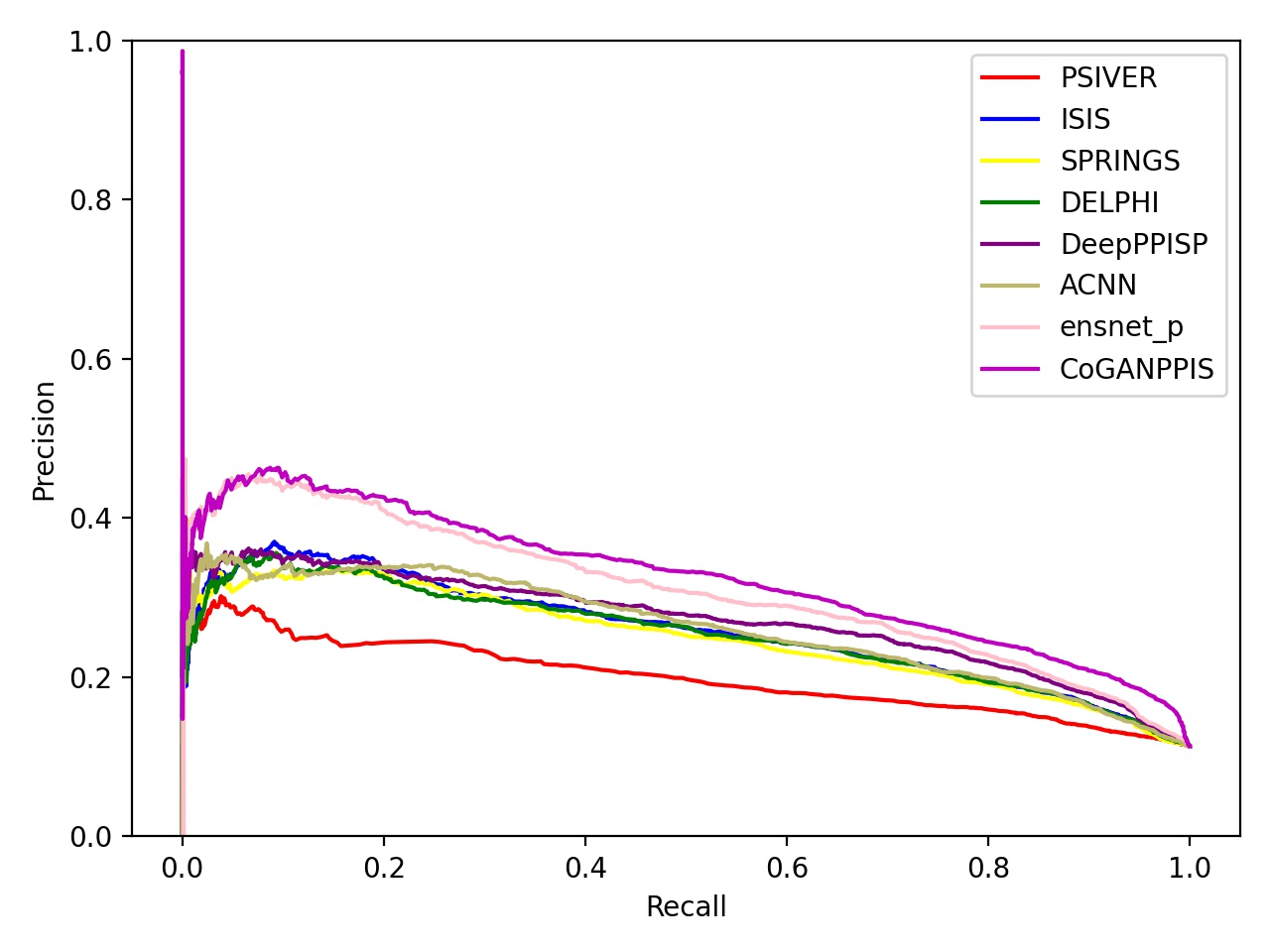}}
    \caption{ROC and PR plots of CoGANPPIS and other seven competing models. (a) ROC plot on Dset422, (b) PR plot on Dset422, (c) ROC plot on Dset448, (d) PR plot on Dset448. CoGANPPIS clearly outperforms other models}
	\label{ROC_PR}
\end{figure}

Table \ref{performance} presents the experimental results of seven sequence-based competitive PPIs prediction models and our proposed model. It can be observed that CoGANPPIS achieves the best performance across both two datasets in terms of ${\rm F_1}$ and ${\rm MCC}$ consistently, which ascertains its effectiveness. The ROC and PR curves of CoGANPPIS and other competing methods on Dset422 and Dset448 are shown in Figure \ref{ROC_PR}. It demonstrates that CoGANPPIS has higher AUPRC and AUC than other competing methods.

\begin{table*}
\footnotesize
\renewcommand\arraystretch{1.2}
\caption{Performance comparison\label{performance}}
\tabcolsep=0pt
\begin{tabular*}{\textwidth}{@{\extracolsep{\fill}}ccccccccccccc@{\extracolsep{\fill}}}
\toprule%
Dataset & \multicolumn{6}{@{}c@{}}{Dset422} & \multicolumn{6}{@{}c@{}}{Dset448} \\
\cline{2-7}\cline{8-13}%
Method & AUPRC & ACC & Recall & Precision & ${\rm F_1}$ & MCC & AUPRC & ACC & Recall & Precision & ${\rm F_1}$ & MCC\\
\midrule
PSIVER & 0.230 & 0.690 & 0.324 & 0.241 & 0.276 & 0.086 & 0.238 & 0.761 & 0.242 & 0.306 & 0.270 & 0.131 \\
    ISIS & 0.284 & 0.767 & 0.328 & 0.351 & 0.339 & 0.198 & 0.316 & 0.730 & 0.440 & 0.324 & 0.373 & 0.210 \\
    SPRINGS & 0.279 & 0.700 & 0.357 & 0.263 & 0.303 & 0.120 & 0.305 & 0.753 & 0.266 & 0.300 & 0.282 & 0.134 \\
    DELPHI & 0.311 & 0.770 & 0.376 & 0.370 & 0.373 & 0.232 & 0.320 & 0.764 & 0.394 & 0.365 & 0.379 & 0.234 \\
    DeepPPISP & 0.320 & 0.655 & 0.577 & 0.303 & 0.397 & 0.206 & 0.351 & 0.772 & 0.406 & 0.383 & 0.394 & 0.254 \\
    ACNN & 0.306 & \textbf{0.775} & 0.342 & 0.371 & 0.356 & 0.220 & 0.301 & \textbf{0.782} & 0.339 & 0.388 & 0.362 & 0.232 \\
    ensnet\_p & 0.377 & 0.766 & 0.414 & \textbf{0.372} & 0.392 & 0.248 & 0.385 & 0.775 & 0.405 & \textbf{0.389} & 0.397 & 0.259 \\
    CoGANPPIS & \textbf{0.385} & 0.703 & \textbf{0.582} & 0.325 & \textbf{0.418} & \textbf{0.259} & \textbf{0.388} & 0.752 & \textbf{0.490} & 0.366 & \textbf{0.419} & \textbf{0.270} \\
\hline
\end{tabular*}
\end{table*}

\subsection{Ablation analysis}

In this part, we test the usefulness of introducing coevolutionary information into PPIs prediction. First, we evaluate the performance of the model with coevolutionary information included. Then we remove coevolutionary information from the model and train the model again to measure the comparing performance. The model without coevolutionary information is named as ${\rm CoGANPPIS}^{\ominus}$. Table \ref{Ablation_exp} demonstrates the performance of CoGANPPIS and 
${\rm CoGANPPIS}^{\ominus}$. We can find that CoGANPPIS outperforms ${\rm CoGANPPIS}^{\ominus}$ on both two datasets, which indicates that introducing coevolutionary information could help improve predictive accuracy and thus validates its effectiveness in PPIs prediction.

\begin{table*}
\footnotesize
\renewcommand\arraystretch{1.2}
\caption{Ablation analysis\label{Ablation_exp}}
\tabcolsep=0pt
\begin{tabular*}{\textwidth}{@{\extracolsep{\fill}}ccccccccccccc@{\extracolsep{\fill}}}
\toprule%
Dataset & \multicolumn{6}{@{}c@{}}{Dset422} & \multicolumn{6}{@{}c@{}}{Dset448} \\
\cline{2-7}\cline{8-13}%
Method & AUPRC & ACC & Recall & Precision & ${\rm F_1}$ & MCC & AUPRC & ACC & Recall & Precision & ${\rm F_1}$ & MCC\\
\midrule
${\rm CoGANPPIS}^{\ominus}$ & 0.373 & \textbf{0.758} & 0.449 & \textbf{0.366} & 0.403 & 0.255 & 0.374 & \textbf{0.763} & 0.443 & \textbf{0.375} & 0.406 & 0.262 \\
CoGANPPIS & \textbf{0.385} & 0.703 & \textbf{0.582} & 0.325 & \textbf{0.418} & \textbf{0.259} & \textbf{0.388} & 0.752 & \textbf{0.490} & 0.366 & \textbf{0.419} & \textbf{0.270} \\
\hline
\end{tabular*}
\end{table*}

\subsection{Model performance on proteins of different lengths}

\begin{figure}[H]
	\centering  
	\subfigbottomskip=0pt 
	\subfigcapskip=0pt 
	\subfigure[]{
	    \label{2a}
		\includegraphics[width=0.24\linewidth]{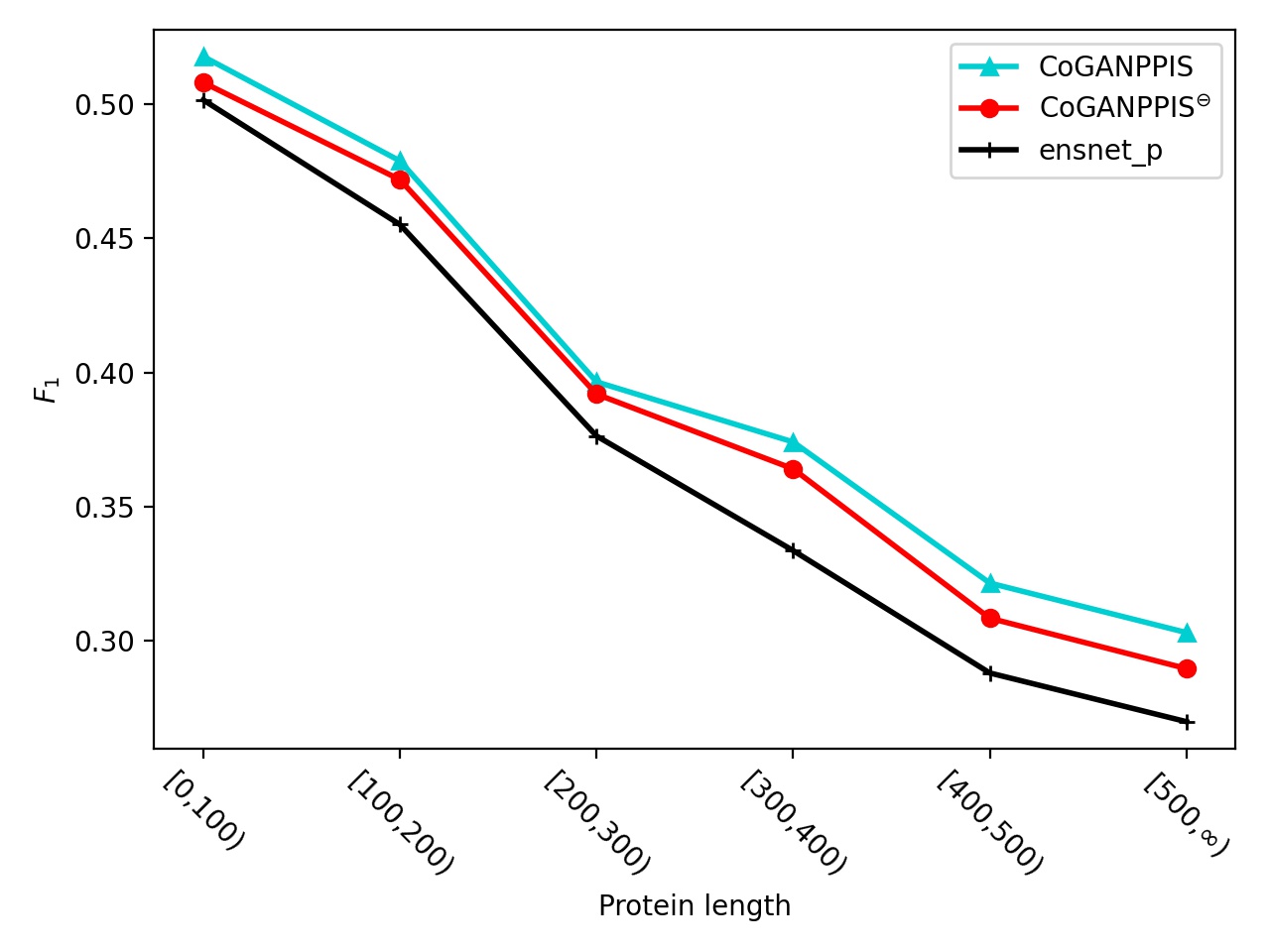}}
	\subfigure[]{
	    \label{2b}
		\includegraphics[width=0.24\linewidth]{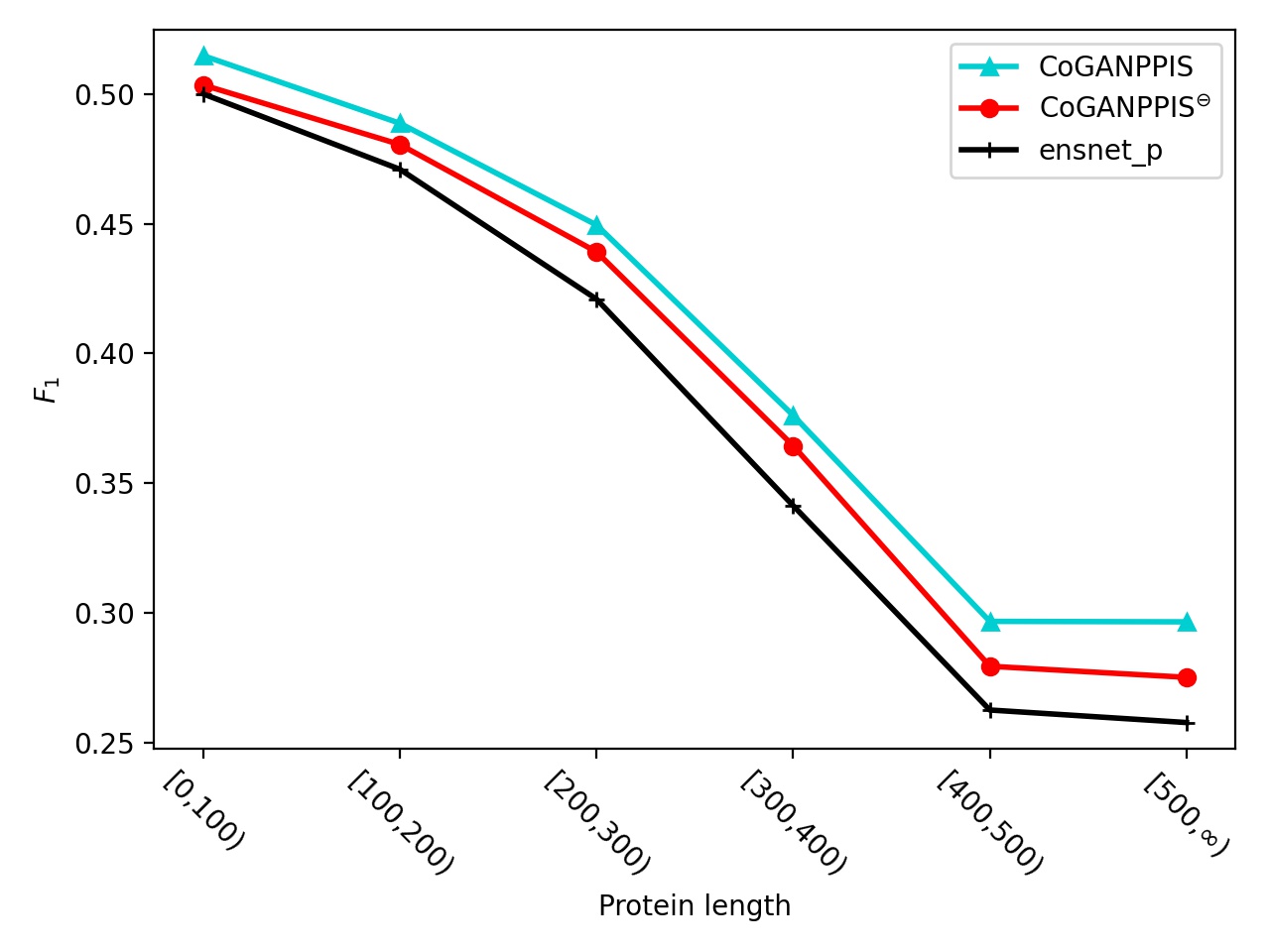}}
	  \\
	\subfigure[]{
	    \label{2c}
		\includegraphics[width=0.24\linewidth]{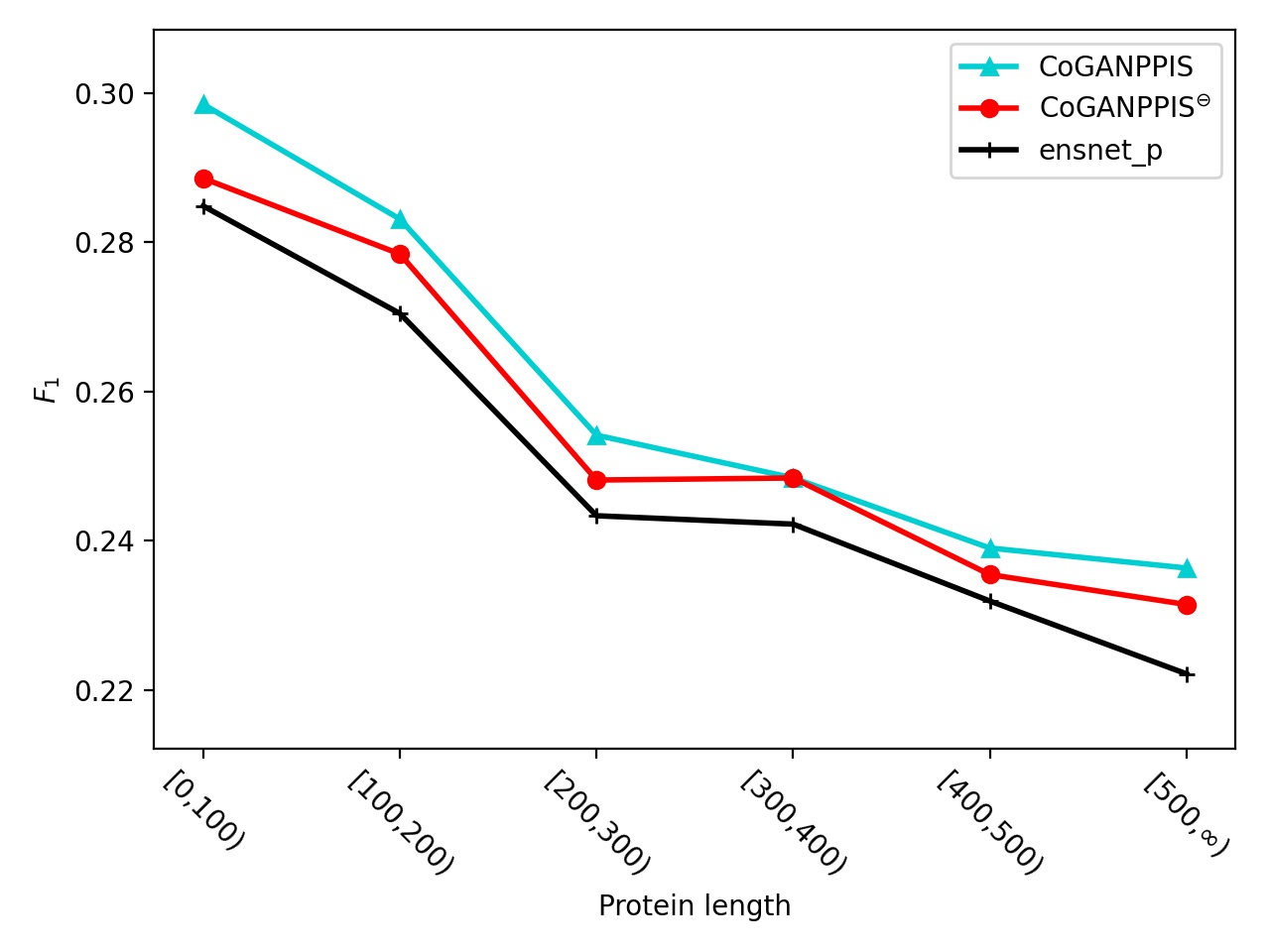}}
	\subfigure[]{
	    \label{2d}
		\includegraphics[width=0.24\linewidth]{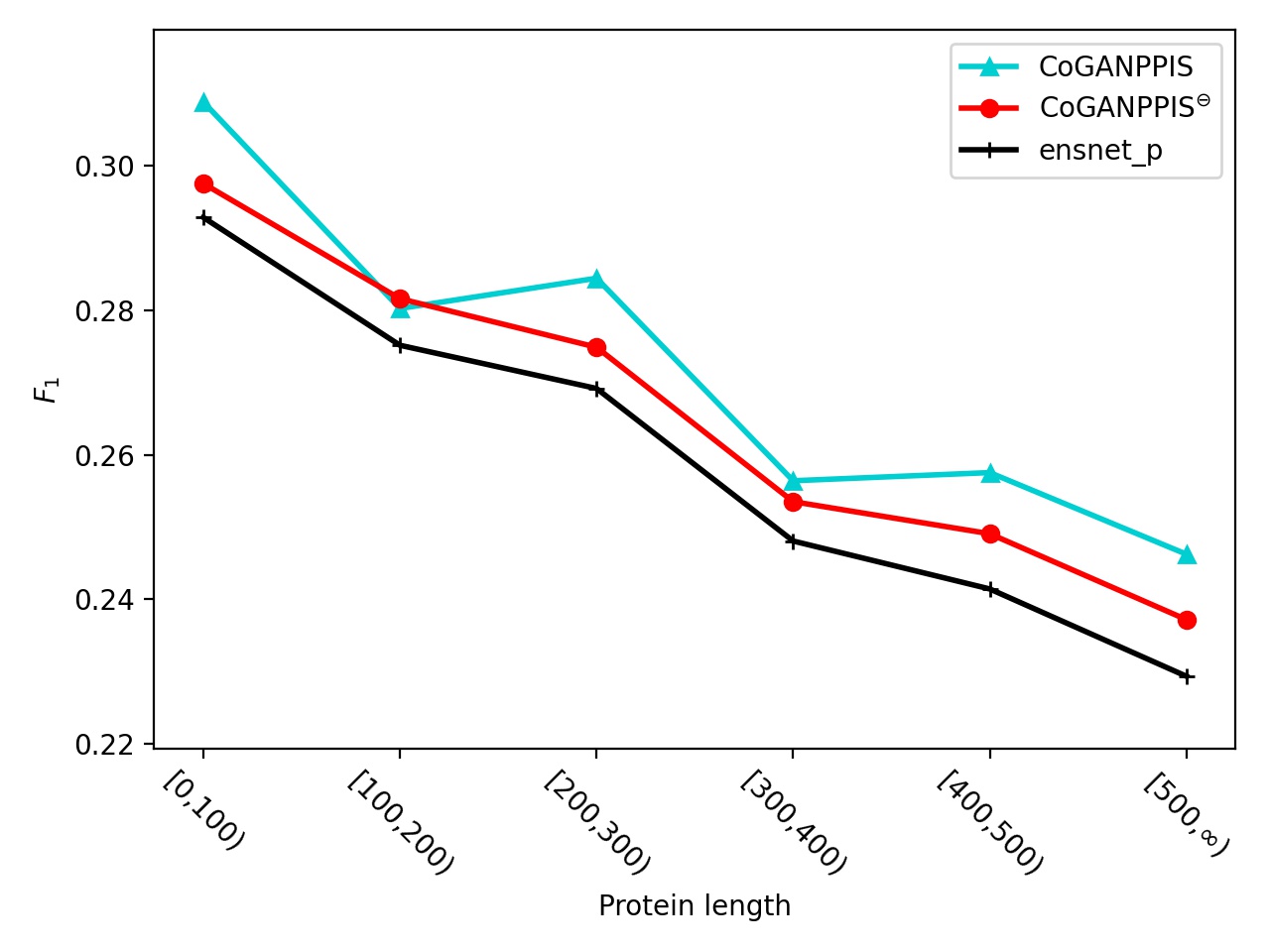}}
	\caption{Models' performance under various protein lengths. (a) $F_1$ on Dset422, (b) $F_1$ on Dset448, (c) MCC on Dset422, (d) MCC on Dset448}
	\label{F_MCC_on_diff_protein_length}
\end{figure}

Considering that protein sequence length varies greatly from each other, it could be necessary to study the predictive performance on proteins of different lengths. To answer this question, we plot the experimental results of CoGANPPIS, ${\rm CoGANPPIS}^{\ominus}$ as well as ensnet\_p in terms of $F_1$ and MCC under proteins of different lengths on the two datasets in Figure \ref{F_MCC_on_diff_protein_length}.

Through the results, we have the following observations: First, we can observe that with the protein sequence length increasing, the performance of three models on two datasets all show an overall downward trend. This can be explained as the protein structure and function become more complex with the increase of length, making PPIs more difficult to predict. Second, it is interesting that the performance improvement of CoGANPPIS and ${\rm CoGANPPIS}^{\ominus}$ compared with ensnet\_p increases as the increase of the protein sequence length. Take the $F_1$ on Dset422 in Figure \ref{2a} as an example, when the length is less than 100, the $F_1$ of CoGANPPIS and ${\rm CoGANPPIS}^{\ominus}$ are 0.518 and 0.508 respectively, which are only 0.016 and 0.006 higher than that of ensnet\_p (0.502). When the protein length is between 200 and 300, the improvements increase to 0.020 and 0.016 (0.396 vs 0.376 and 0.392 vs 0.376). When the protein length is greater than 500, the gaps further increase to 0.033 and 0.021 (0.302 vs 0.269 and 0.290 vs 0.269). This clearly shows that the longer the protein sequence, the more PPIs prediction relies on global information extraction, which can be better captured by our global attention mechanism (even without coevolutionary information). Third, we pay attention to the comparison between CoGANPPIS and ${\rm CoGANPPIS}^{\ominus}$. The two metrics $F_1$ and MCC of CoGANPPIS on both two datasets are better than the ones of ${\rm CoGANPPIS}^{\ominus}$, which also verifies the effectiveness of coevolutionary information in PPIs prediction and confirms the conclusion that we obtained in ablation analysis.

\begin{figure}[H]
	\centering  
	\subfigbottomskip=0pt 
	\subfigcapskip=0pt 
	\subfigure[]{
	    \label{trend}
		\includegraphics[width=0.24\linewidth]{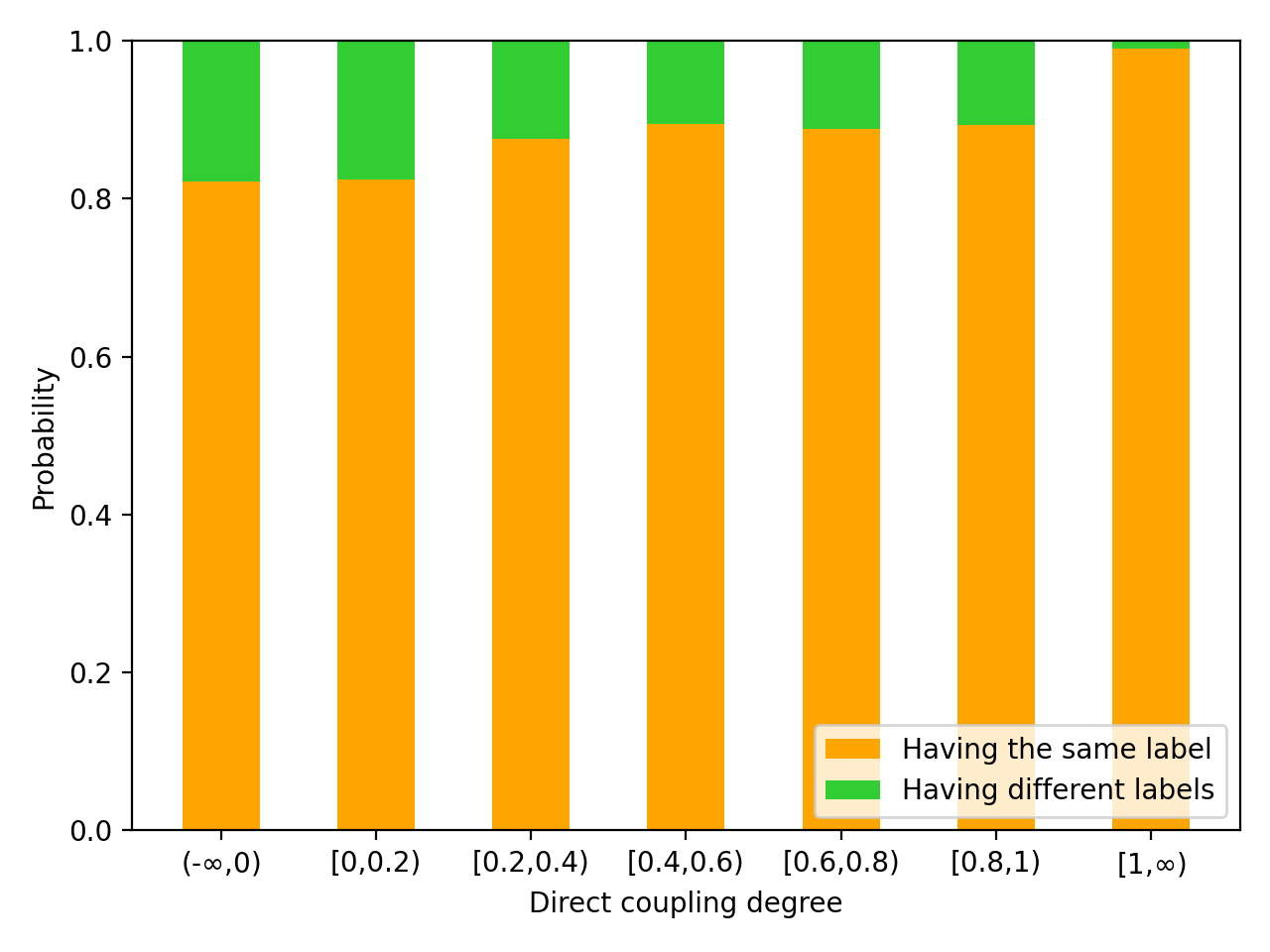}}
	\subfigure[]{
	    \label{scatter_dca}
		\includegraphics[width=0.24\linewidth]{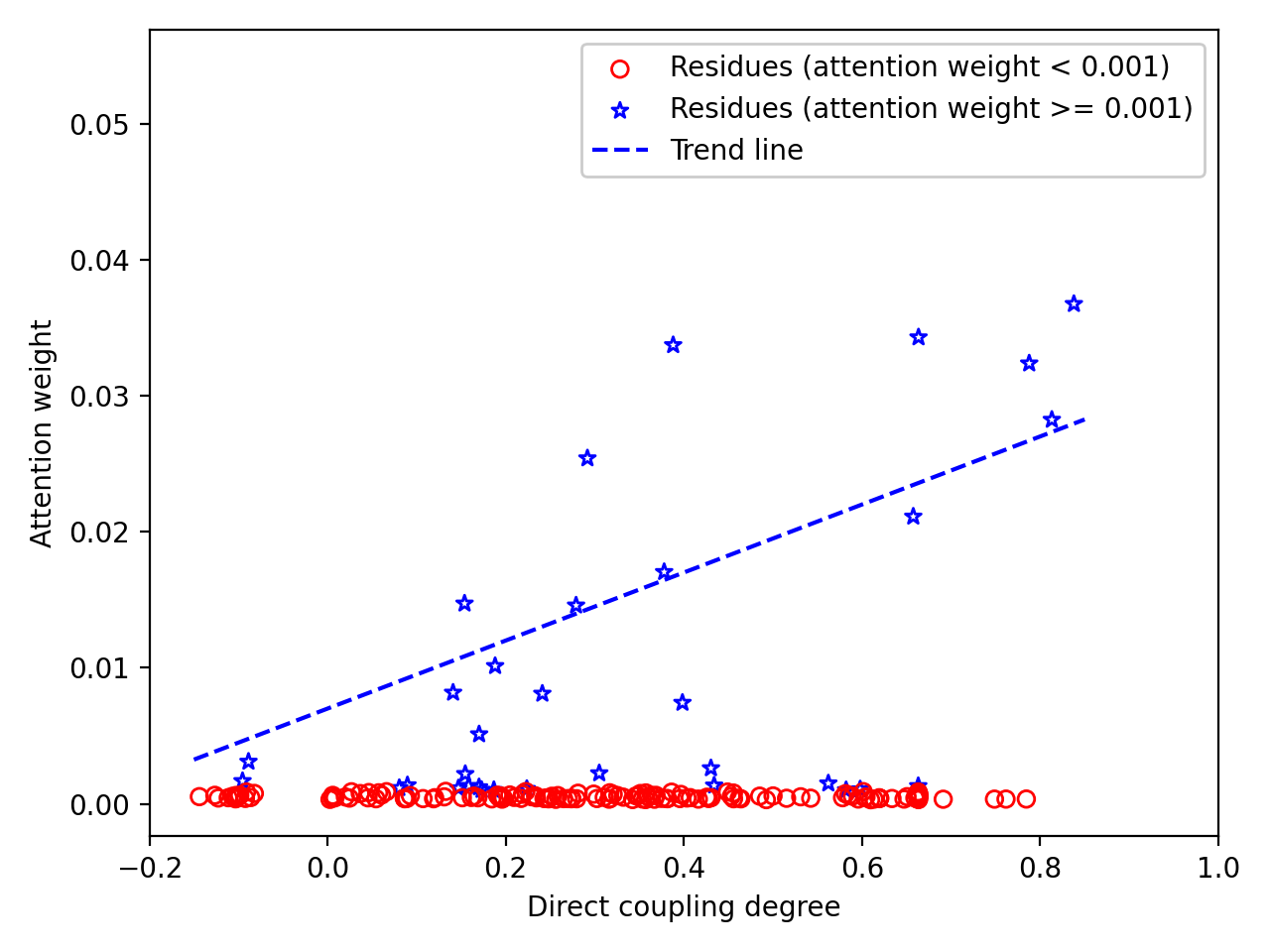}}
	\caption{Analysis of coevolution-enhanced global attention mechanism. (a) We plot the probabilities of residue pairs having the same label (orange) and the probabilities of having different labels (green) under different direct coupling degrees in our dataset, (b) We take protein Q2T3W4 as an example and plot the scatter diagram of its attention weights under different direct coupling degrees. We also paint the trend line of the scattered points with attention weights larger than 0.001, which shows the positive correlation between the two variables}
	\label{protein_attention_mechanism}
\end{figure}

\begin{figure*}
	\centering  
	\subfigbottomskip=2pt 
	\subfigcapskip=-5pt 
	\subfigure[Coevolutionary attention]{
	    \label{coattent_dis}
		\includegraphics[width=0.27\linewidth]{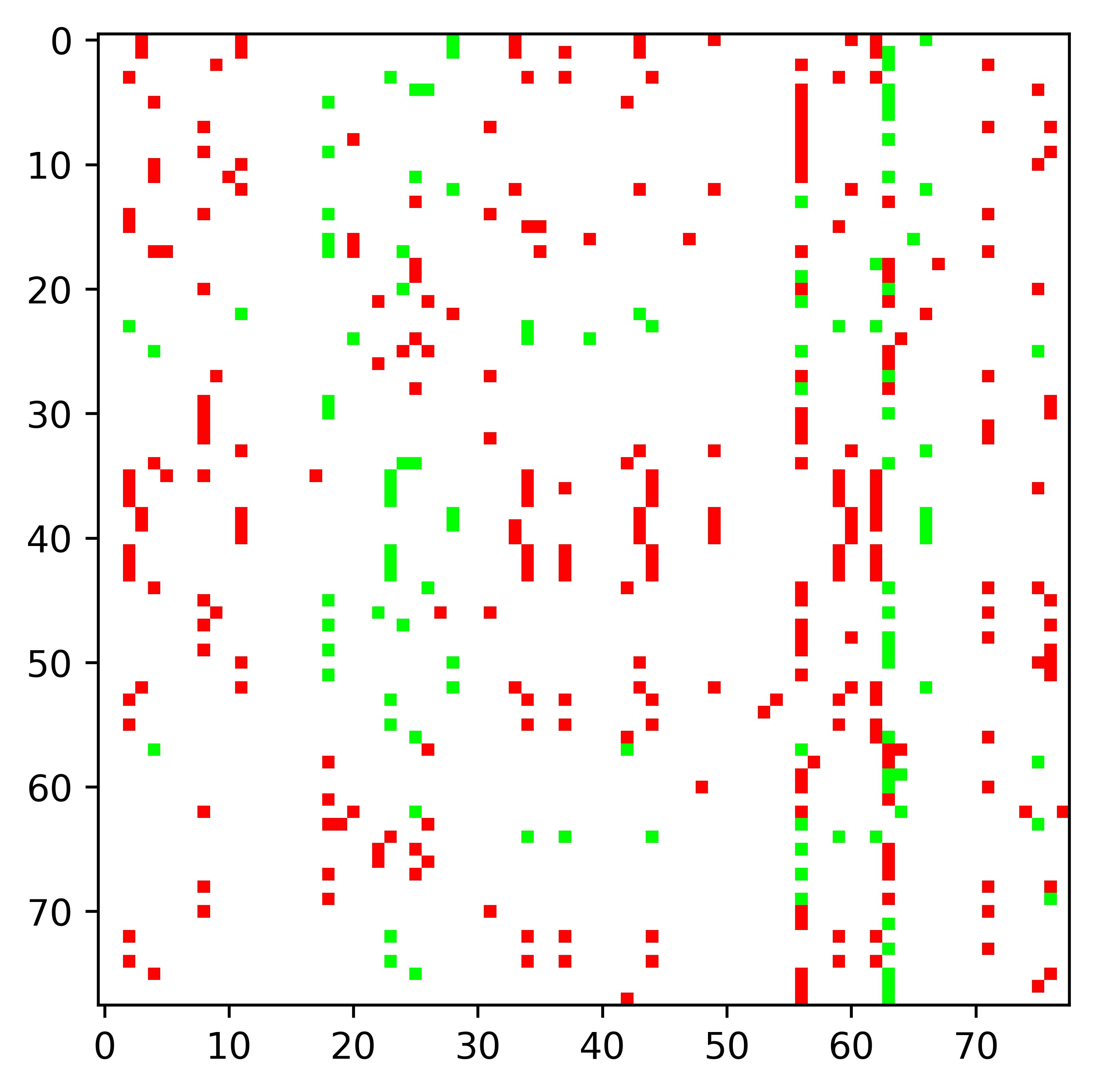}}
	\subfigure[DCA]{
	    \label{dca_dis}
		\includegraphics[width=0.27\linewidth]{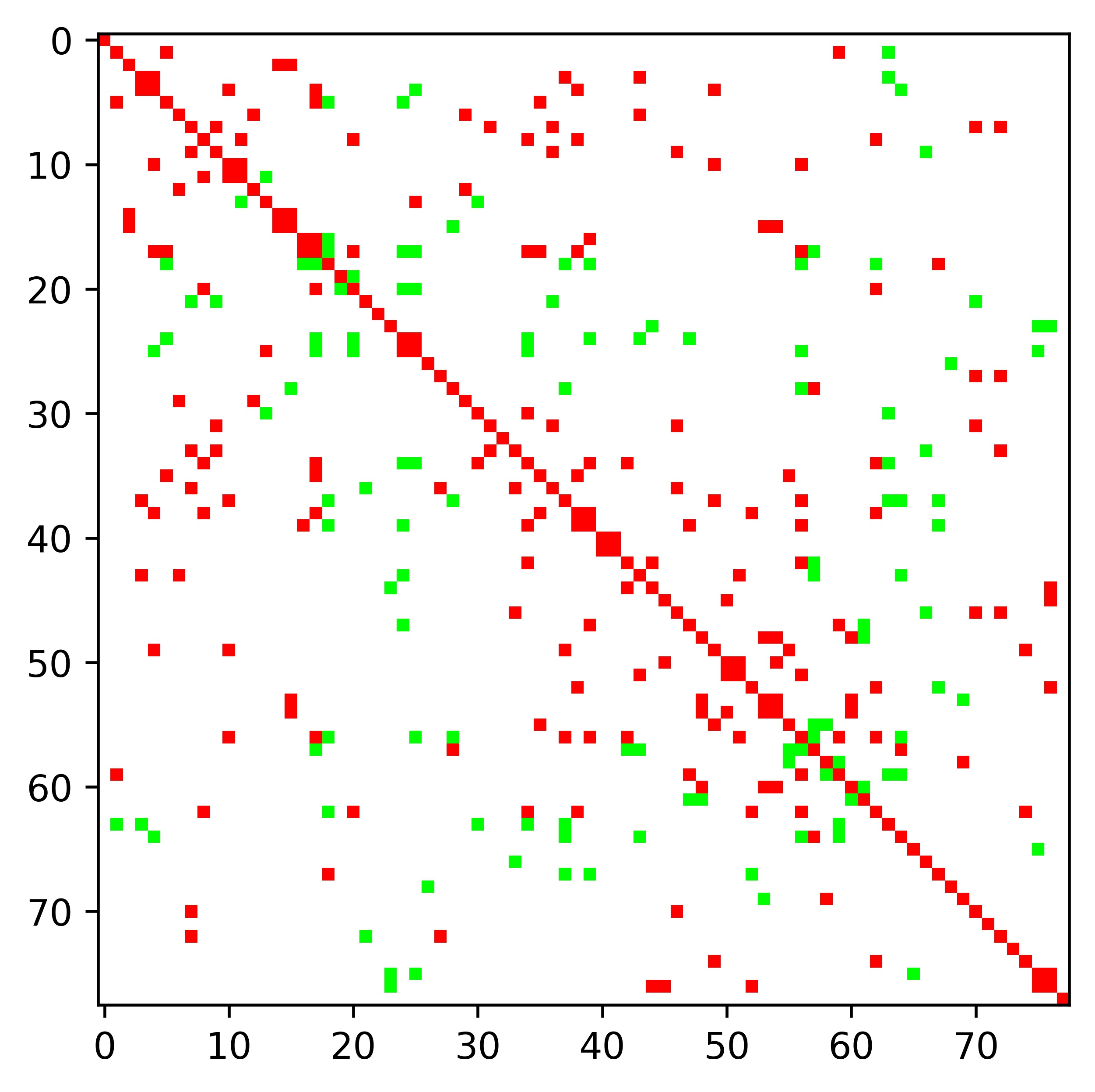}}
	\subfigure[Comparison on different separations]{
	    \label{sparation}
		\includegraphics[width=0.35\linewidth]{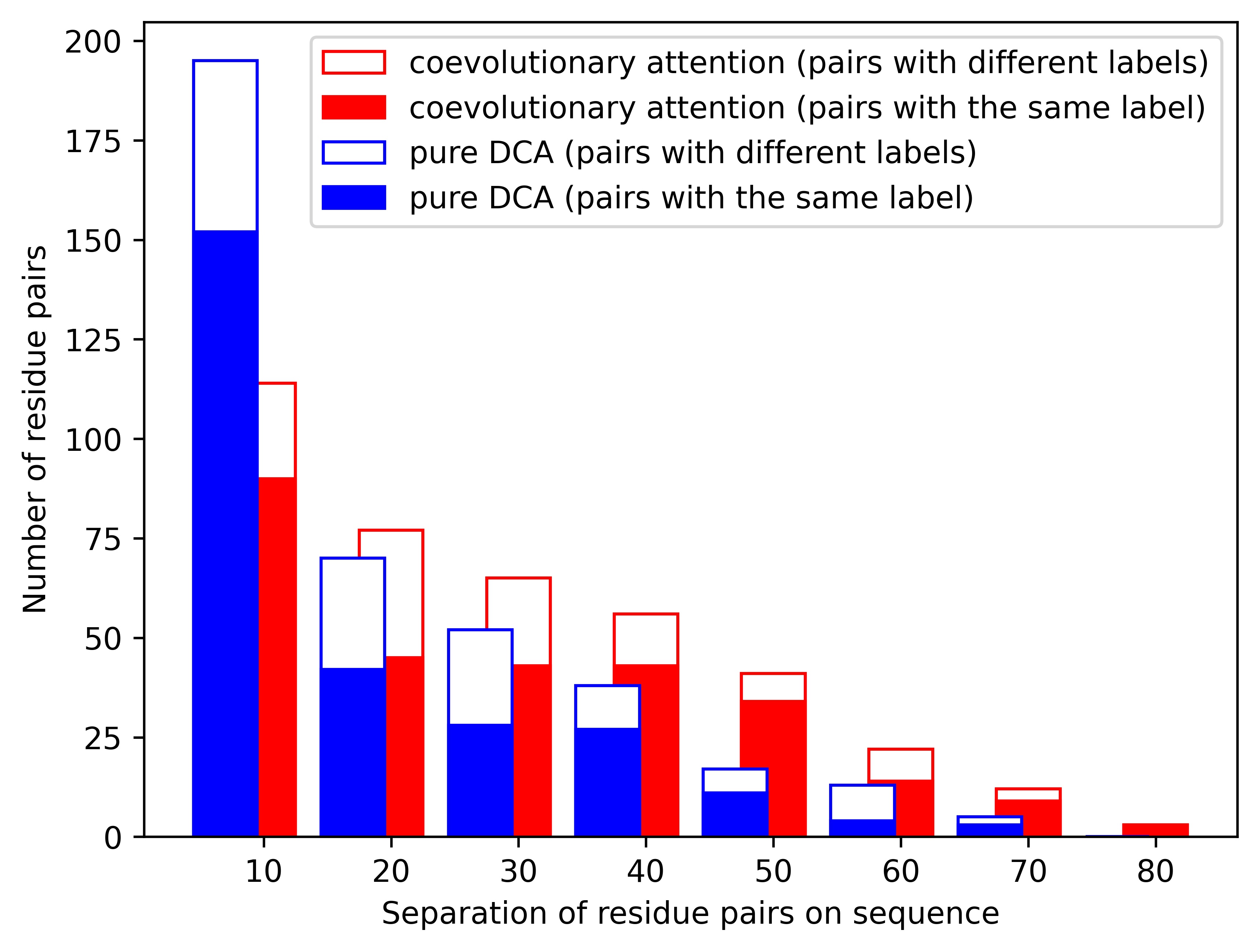}}
	
	\subfigure[Coevolutionary attention]{
	    \label{coattent_dis}
		\includegraphics[width=0.27\linewidth]{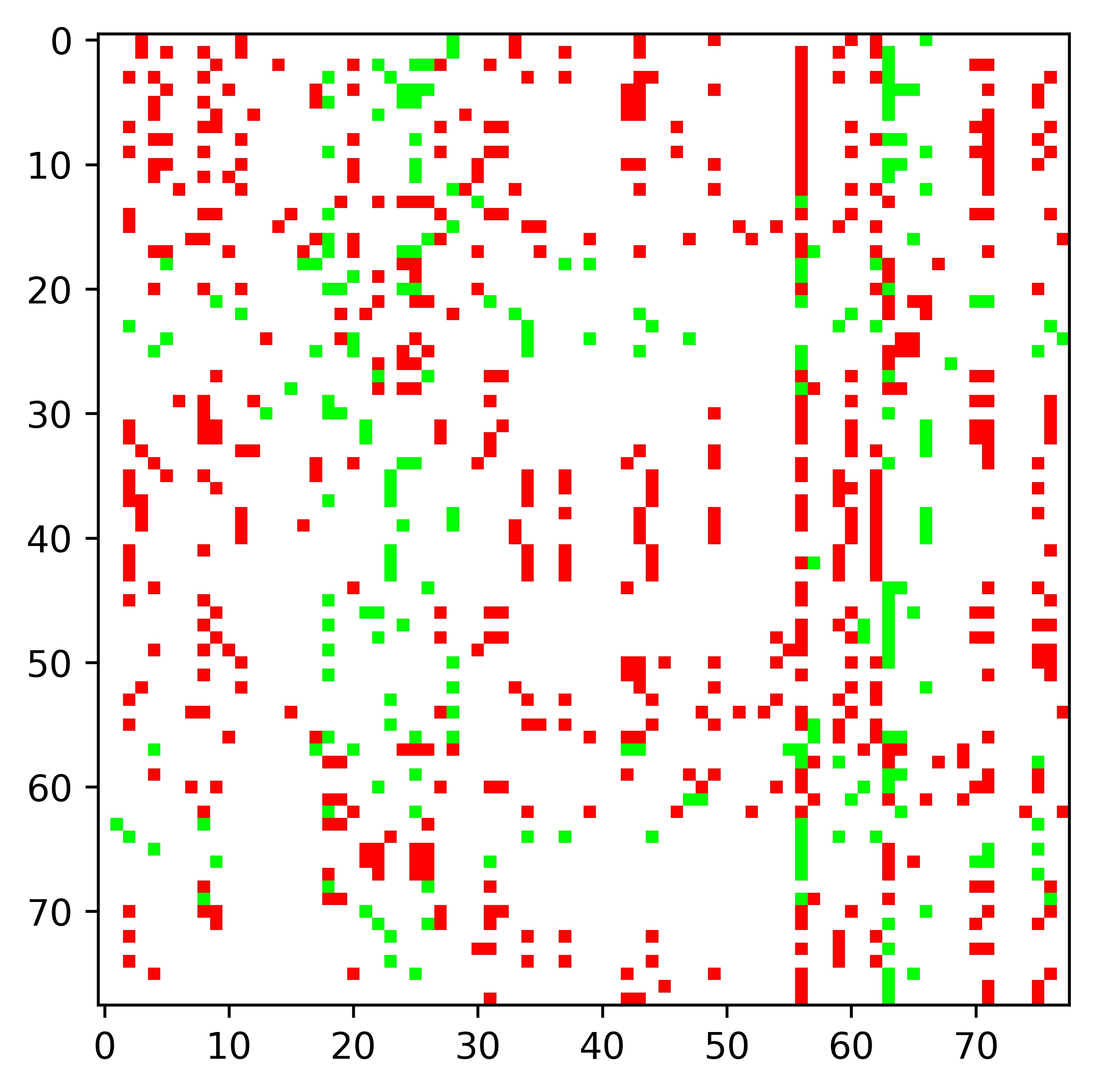}}
	\subfigure[DCA]{
	    \label{dca_dis}
		\includegraphics[width=0.27\linewidth]{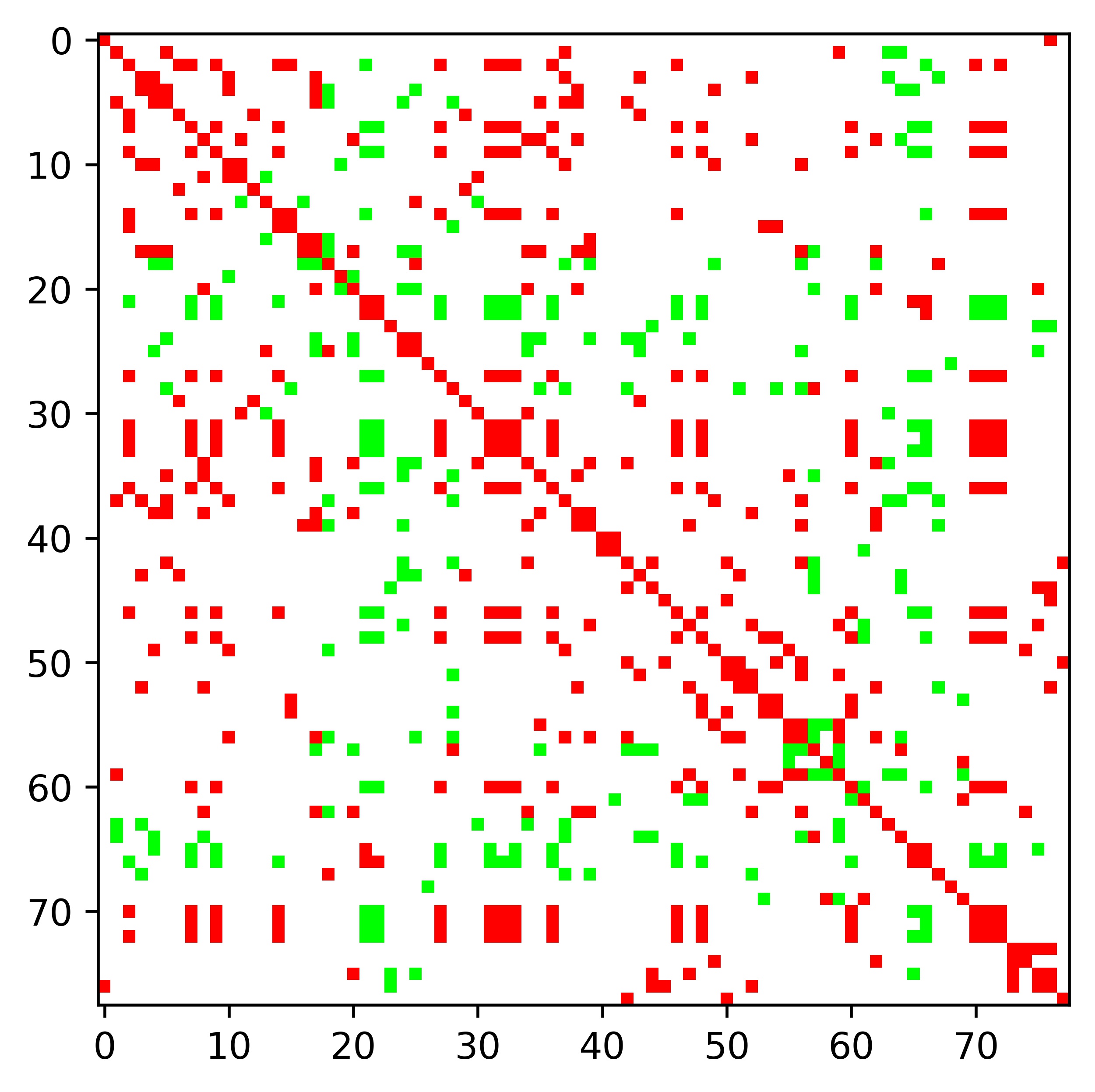}}
	\subfigure[Comparison on different separations]{
	    \label{sparation}
		\includegraphics[width=0.35\linewidth]{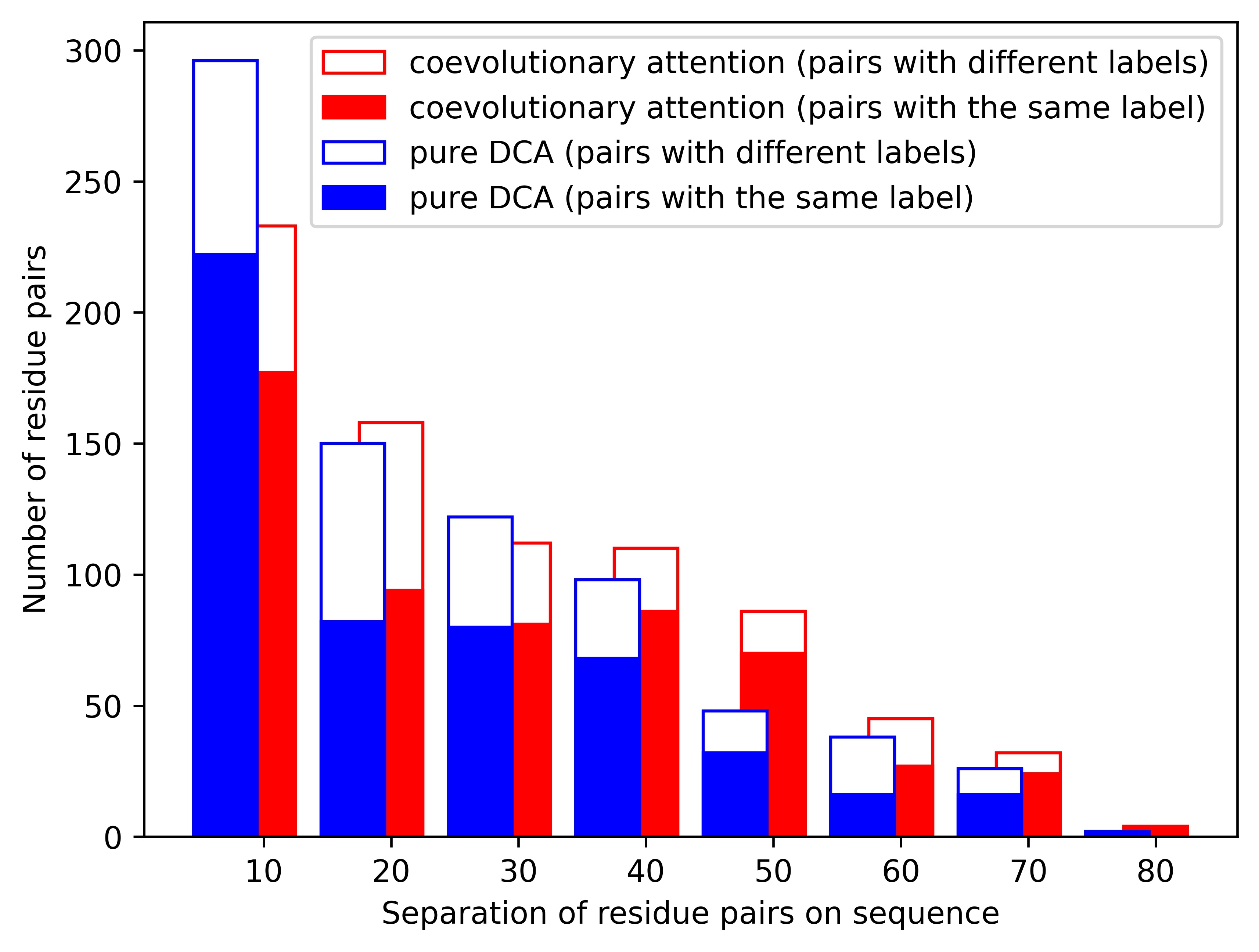}}
	
	\subfigure[Coevolutionary attention]{
	    \label{coattent_dis}
		\includegraphics[width=0.27\linewidth]{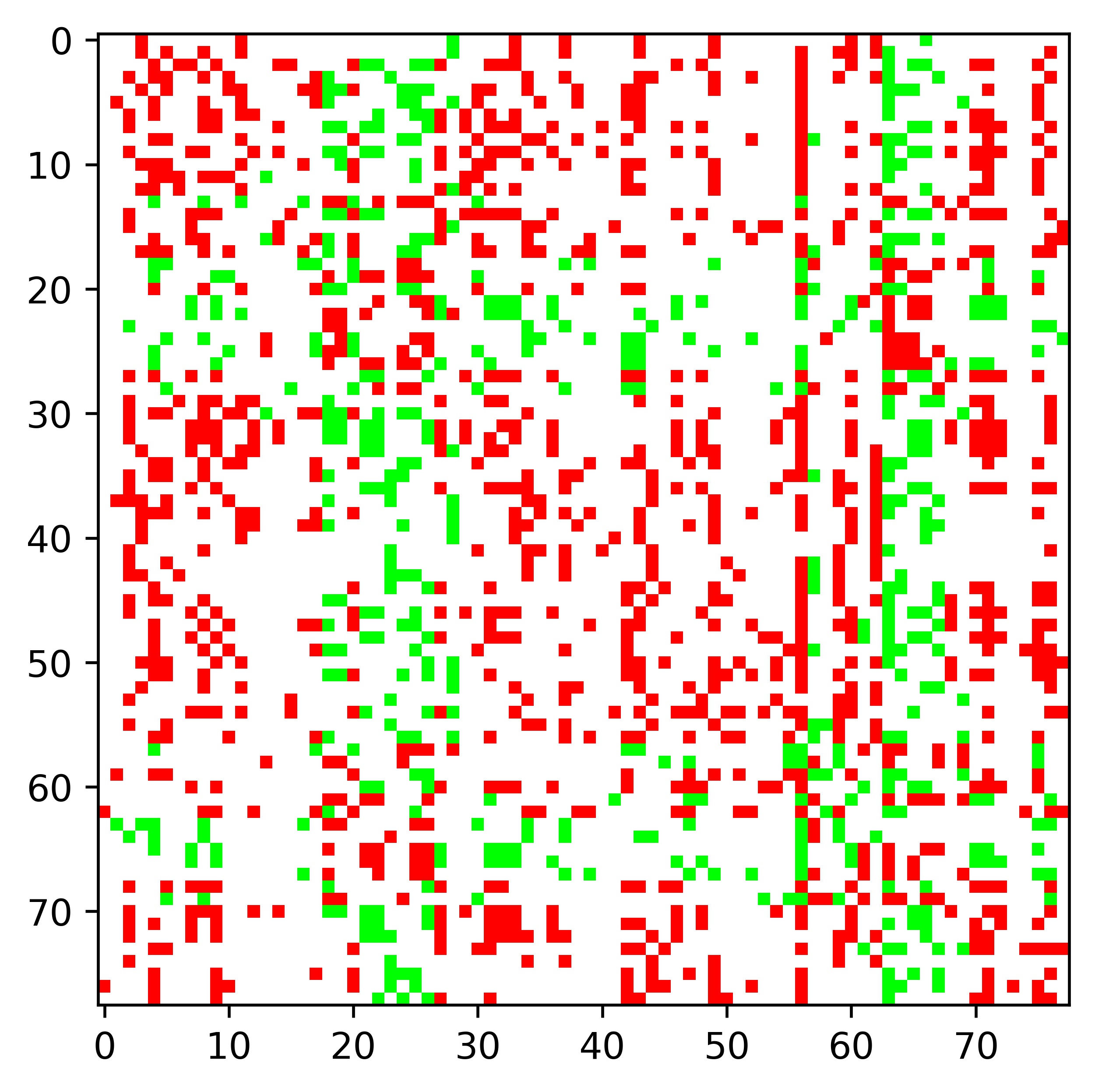}}
	\subfigure[DCA]{
	    \label{dca_dis}
		\includegraphics[width=0.27\linewidth]{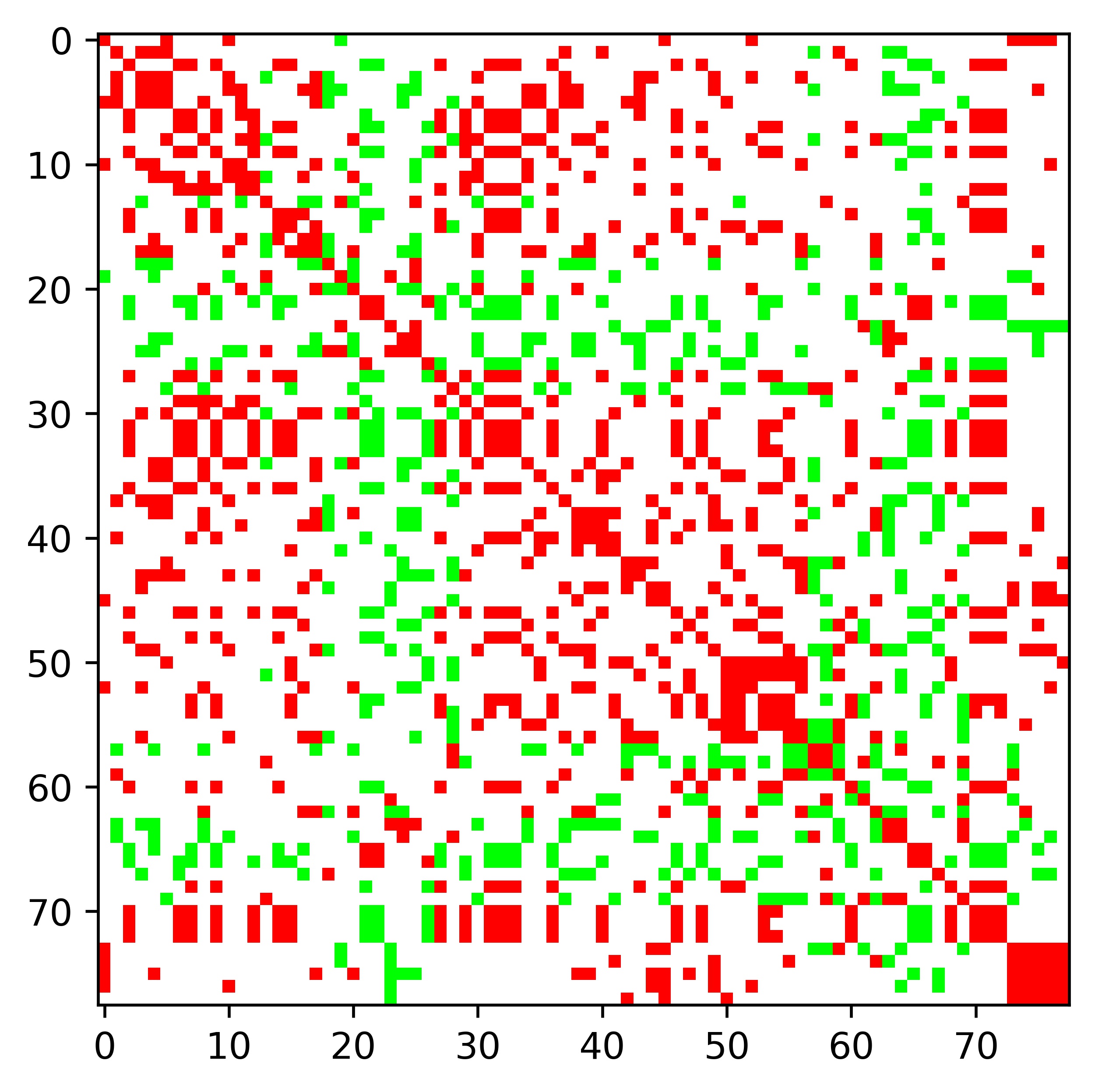}}
	\subfigure[Comparison on different separations]{
	    \label{sparation}
		\includegraphics[width=0.35\linewidth]{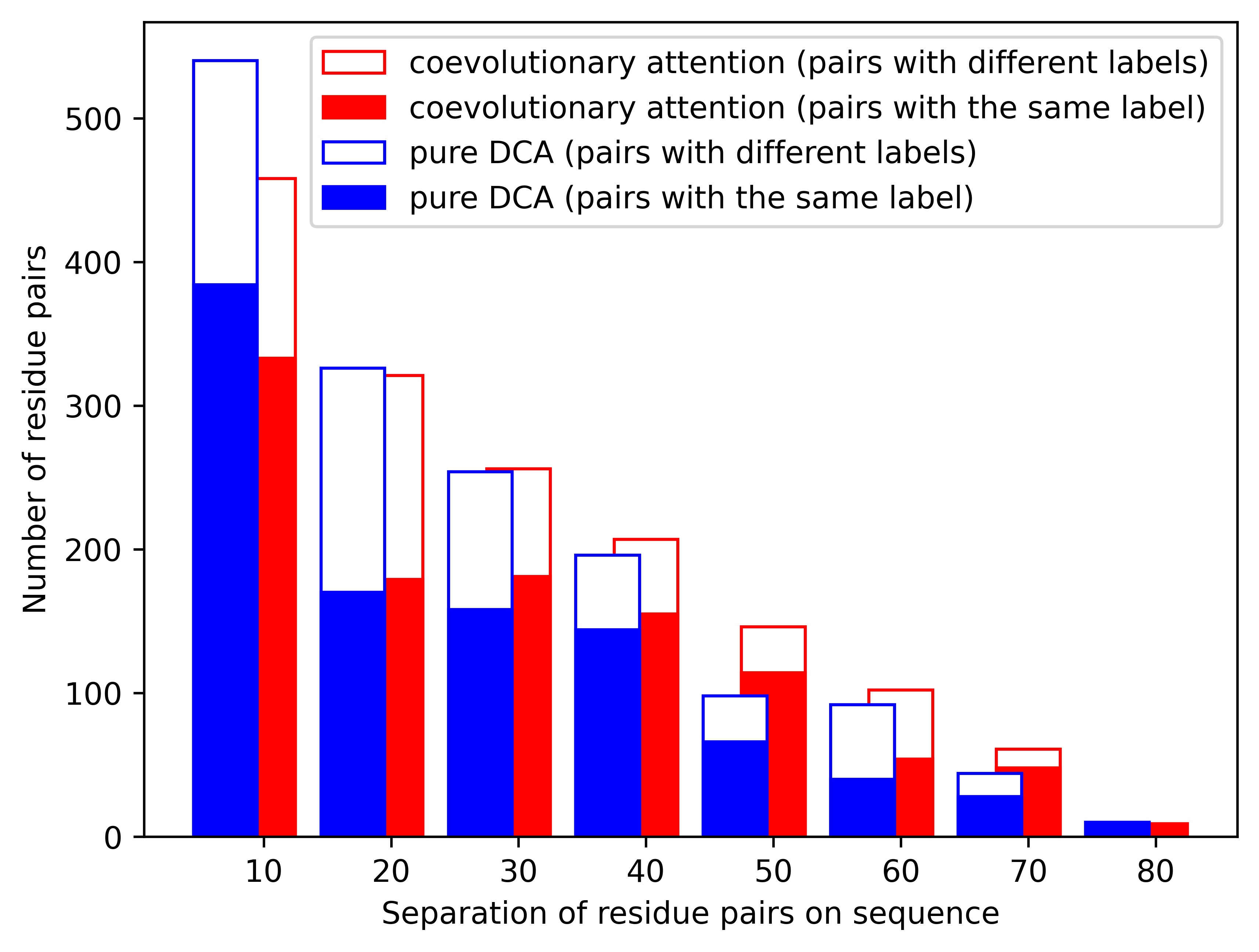}}
	\caption{Visualization of attention weights of protein 2fcw\_B using our attention mechanism (a, d and g) and pure DCA (b, e and h). Residue pairs are sorted according to the attention weights and the direct coupling degrees respectively, and kept the 5N (a-c), 10N (d-f) as well as 20N (g-i) highest-ranking pairs. These pairs with same labels are indicated in red and the pairs with different labels are shown in green. The right-most panels (c, f and i) bin weights of residue pairs according to their separation along the protein sequence. Note that our attention mechanism captures the residues with same labels more accurately than pure DCA. Besides, the residues paid attention to by our attention mechanism are more evenly distributed than pure DCA}\label{protein_attention_mechanism_dist}
\end{figure*}

\subsection{Analysis of coevolution-enhanced global attention mechanism}

We have verified the effectiveness of coevolutionary information and global attention mechanism. Now let's further study why the coevolution-enhanced global attention mechanism works. First, for each pair of residues, we examine the relationships between its direct coupling degrees and labels. As shown in Figure \ref{trend}, the larger the direct coupling degree, the higher the probability that the pairs of residues have the same label. Further, let's take protein Q2T3W4 in Dset448 as an example. We first extract the attention weights of the first residue on Q2T3W4 in the training process. Then we plot a scatter diagram and fit a trend line of the points with attention weights greater than 0.001 as shown in Figure \ref{scatter_dca}. We find that the slope of the trend line is positive, which implies that in general, the larger the direct coupling degrees, the higher the attention weights. Hence, the target residue could pay more attention to the residues with high correlation during training process, which is an noticeable advantage of this attention mechanism.

\subsection{Visualization}

As mentioned above, the attention weights between two residues show a positive correlation with their direct coupling degrees. It prompts us to explore to which extent our attention mechanism enhances compared with allocating weights purely according to DCA. For a protein sequence with $N$ amino acid residues, we sort residue pairs according to their attention weights and direct coupling degrees respectively, and keep the $5N$, $10N$ as well as $20N$ highest-ranking pairs. Figure \ref{protein_attention_mechanism_dist} shows an example of the protein 2fcw\_B from Dset422. The three rows refer to the $5N$, $10N$ and $20N$ situations in turn (Only the corresponding number of selected residue pairs are colored). The pairs with same labels are indicated in red and the pairs with different labels are shown in green. It is evident that our attention mechanism works better than pure DCA because of its higher proportion of red points in all three situations. To become more quantitative, we have binned the selected pairs according to their separation along the protein sequence as shown in the third column in Figure \ref{protein_attention_mechanism_dist}. We observe that our attention mechanism captures the residues with same labels more accurately than pure DCA. Also, our attention mechanism could attach more attention weights to distant residues, whereas pure DCA tends to pay more attention to neighboring residues, which could be attributed to the consideration of the whole protein feature representation of our attention mechanism.




%
%

\section{Conclusion}

The aim of this paper is to improve the PPIs prediction performance solely based on protein sequences, which is important for understanding the biological mechanism of proteins both experimentally and theoretically. A dozen of sequence-based PPIs predictors have been developed in recent years. However, most of these works just utilize some commonly used features without considering coevolutionary information which provides rich clues for inter-residue relationships. Also, they are not good at predicting PPIs of long-length proteins.

Here, we propose a coevolution-enhanced global attention neural network (CoGANPPIS). Specifically, we employ a coevolution-enhanced global attention mechanism both for better inter-residue relationship capture and for better understanding of long-length proteins. We further aggregate the local residue features and apply a CNN \& pooling layer to the coevolutionary information features as a supplement. Then we utilize several fully connected layers to generate the final prediction. Extensive experiments of CoGANPPIS and seven other popular methods on two standardized datasets show that our proposed model CoGANPPIS achieves the state-of-the-art performance.

Further experimental analysis shows that: (1) Coevolutionary information can improve the performance of PPIs prediction. (2) CoGANPPIS can bring more performance improvement compared with previous methods as the protein sequence becomes longer, implying that CoGANPPIS has a better understanding of the whole protein sequences. (3) Compared with allocating attention weights according to pure DCA, the proposed coevolution-enhanced global attention mechanism pays more attention to the residues with same labels and displays a more evenly distributed attention weights instead of locally aggregated attention weights.

Although CoGANPPIS shows advantages over previous methods, it has some limitations: First, CoGANPPIS takes a lot of computation time due to its usage of multiple sequence alignments and direct coupling analysis to generate coevolutionary information. In addition, DCA's accuracy depends on the number of homologs, making it difficult for those proteins with little homologs. In the future, we would be commited to find useful, practical and convenient features to make prediction faster and more accurate.



\section*{Funding}

This study was supported by funding from the National Natural Science Foundation of China (72222009, 71991472 to X.Z), the National Natural Science Foundation of China (3210040426 to Z.H.), the Shanghai Rising-Star Program (21QB1400900 to Z.H.), and was also partly supported by a grant from the major project of Study on Pathogenesis and Epidemic Prevention Technology System (2021YFC2302500) by the Ministry of Science and Technology of China.\vspace*{-12pt}

\bibliography{reference23}
\end{document}